\documentclass[11pt]{article}

\usepackage[latin1]{inputenc}
\usepackage{graphicx}
\usepackage{epsfig}
\newenvironment{proof}{\trivlist
\item[\hskip\labelsep{\it Proof}\,:]}{\hfill{$q.e.d$}\endtrivlist}

\newtheorem{theorem}{Theorem}[section]
\newtheorem{lemma}{Lemma}[section]
\newtheorem{proposition}{Proposition}[section]
\newtheorem{corollary}{Corollary}[section]

\newtheorem{remark}{Remark}[section]

\newfont{\bb}{msbm10 at 10pt}
\def\r{\hbox{\bb R}}
\def\l{\hbox{\bb L}}

\begin{document}

\title{Stationary liquid drops in Lorentz-Minkowski  space}
\author{Rafael L\'opez\\
Departmento de Geometría y Topología\\
Universidad de Granada\\
18071 Granada (Spain)\\
e-mail:{\tt rcamino@ugr.es}}

\date{}

\maketitle

\begin{abstract} This paper analyzes the configurations of shapes that shows a spacelike liquid 
 drop in Minkowski space deposited  over a spacelike plane $\Pi$. We assume the presence of a uniform 
gravity field directed toward $\Pi$ and that the volume 
of the drop is prescribed. Our interest are the liquid drops that 
 are critical points of the energy of the corresponding mechanical system and we 
will say then that the liquid drop is stationary. 
In such case, the liquid-air interface is 
determined by the condition that the mean curvature  is a linear function of distance from 
$\Pi$ and that the drop makes a  constant hyperbolic angle of contact 
with the plate $\Pi$. As first result, we shall prove that the liquid drop must be  rotational symmetric with respect to an 
axis orthogonal to $\Pi$.  Then we prove  the 
existence and uniqueness of symmetric solutions for a given angle  of contact with $\Pi$. 
Finally, we shall study the shapes that a liquid drop can adopt  in terms of its size. So, we shall derive estimates of 
its  height, volume and area of the wetted surface on $\Pi$.

\end{abstract}

%%%%%%%%%%%%%%%%%%%%%%%%%%%%%%
\section{Introduction and statement of the main results}
%%%%%%%%%%%%%%%%%%%%%%%%%%%%%%

In  Lorentz-Minkowski three-dimensional space $\l^3$ we 
are interested for  the following 
physical setting. Consider a  liquid drop $X$ of a
prescribed volume  that is adjacent to a solid  surface $\Sigma$, called 
the {\it support surface}. We assume that no chemical reaction occurs between the two materials and these ones
are  homogeneous. We admit the existence  of a uniform gravity  vector  field pointing toward $\Sigma$. 
The energy of the physical system involves the area of the region of contact of $X$
 with $\Sigma$, (the liquid-solid interface) and the surface area of the drop (the liquid-air interface). For this reason, 
we imposes in  $X$ and $\Sigma$  a {\it spacelike causal} condition that allows to consider areas of surfaces. This leads 
to that the gravity is a  timelike vector field.
We seek the configurations that adopt the liquid drop in a state of equilibrium, that is, when the 
energy of the physical system  is critical under perturbations of the system that do not change the amount of liquid of $X$.
The parameters that are determined physically can be: i) the support surface and the angle of contact; ii) the 
area of the region that wets the drop and the angle of contact, or iii) the volume of the drop and the angle of contact with the support surface.

From the mathematical viewpoint, we are studying  the possible shapes of 
a spacelike surface in Minkowski space $\l^3$ whose mean curvature is a  function of its position 
in space, and which meets a given spacelike surface  in a prescribed hyperbolic angle. 
The interior of the liquid drop is a bounded domain $X$ of $\l^3$  
 whose boundary $\partial X$ decomposes into $\partial X={\cal S}\cup \Omega$, 
where ${\cal S}$ is the liquid-air interface and $\Omega=\overline{X}\cap\Sigma$ is the region in $\Sigma$ 
occupied by the part of the drop that wets on $\Sigma$. According to the principle of virtual work, and 
when the equilibrium of the system is achieved, the interface ${\cal S}$ is characterized by the equation 
$\Delta x=2 H  N$
for the position vector $x$ on the free surface ${\cal S}$; here $\Delta$ denotes the Laplace-Beltrami
operator on ${\cal S}$, $N$ is a unit  timelike vector field normal to ${\cal S}$ and  $H$ is  the mean curvature of ${\cal S}$. 
The hyperbolic angle $\beta$ with which ${\cal S}$ and $\Sigma$ intersect along 
$\partial {\cal S}={\cal S}\cap\Sigma$ is determined as a physical constant depending only on the materials. 
Physically, the boundary $\partial {\cal S}$ corresponds with the liquid-air-solid interface of the system. 
When $X$ achieves a state of equilibrium, we shall 
say then that $X$ is a {\it stationary liquid drop}.

 In a vertical gravity vector field, the Euler-Lagrange equation for the drop interface requires that 
 $H$ be a linear function of the height, that is, 
$$H(x)=\kappa x_3(x)+\lambda,\hspace*{1cm}x\in {\cal S},$$
where $x_3$ indicates the third space coordinate.
 Here $\kappa$ is the {\it capillarity constant}
 and, as the angle $\beta$,  depends on the materials involved. The constant $\lambda$ is a Lagrange multiplier arising from the volume constraint. On the other hand, when we talk of contact angle, it is implicitly assumed that the boundary regularity of ${\cal S}$ is enough to ensure that the idea of a normal to ${\cal S}$ at every boundary point  makes sense. For this, we will require ${\cal S}$ to be a sufficiently smooth surface up to the boundary $\partial {\cal S}$.

When $\kappa=0$, no gravity appears in the physical system. In this situation, the liquid drop is modeled by 
a surface ${\cal S}$ with  constant mean curvature $H=\lambda/2$. When the support surface 
is a plane, then the  surface is  a planar disc ($H=0$) or  a hyperbolic cap ($H\not=0$) (see 
 \cite{ap} and Theorem \ref{t1} for another proof). In this sense, our interest is focused for the case that $\kappa\not=0$.

Although in this article we shall consider the case that the support surface is a 
spacelike plane, other interesting cases can appear  for more general geometric configurations,
as for example, a hyperbolic plane, or liquid bridges interconnecting a set of spacelike
planes and hyperbolic planes (see the end of Section \ref{cap-wente}). 

This paper is divided in seven sections and organized as follows. We begin in Section \ref{sec2}  with a preparatory introduction 
where  we give precise definitions and  pose the formulation of  the problem variational for our physical system.
Next, we consider stationary liquid drops resting on a spacelike plane $\Pi$ in 
 a vertical gravity field directed toward $\Pi$. We generalize what it occurs for surfaces with constant mean curvature
and  we shall prove in Section \ref{cap-wente}, Theorem \ref{t1}:

\begin{quote}
{\it Under the effect of the gravity, any stationary liquid drop in Minkowski space $\l^3$  resting in a spacelike plane $\Pi$  
is rotationally symmetric with respect to a straight-line orthogonal to $\Pi$.}
\end{quote}

Assuming then rotational symmetry, the Euler equation becomes an ordinary differential equation for the profile curve
that defines the liquid drop. In successive sections,  we discuss in some detail the solutions of this differential equation starting 
in Section \ref{cap-exi} for the problem of existence and uniqueness. As consequence, we prove (Theorems \ref{sessile-ex} and \ref{pendent-ex}): 

\begin{quote} {\it Let $\kappa$ be a constant of capillarity. 
Given  a spacelike plane $\Pi$ and a real number $\beta$, there exists a stationary  liquid drop
in Minkowski space $\l^3$ supported on $\Pi$ and where $\beta$ is 
 the angle of contact between the drop and $\Pi$. The drop is unique up isometries of 
the ambient space $\l^3$.}
\end{quote}

As in Euclidean space, there exists a qualitative difference of shapes that adopt a drop according to the sign of $\kappa$, that 
is,  {\it sessile} liquid drops when $\kappa>0$ and {\it pendent} liquid drops if $\kappa<0$. In Sections \ref{cap-sessile1} and \ref{cap-sessile2}
we analyze the size of the shapes that a sessile liquid drop can adopt by deriving estimates of 
the height, volume and properties of monotonicity. 
Finally,  Section \ref{cap-pendent} is devoted to the study
of pendent drops. We refer to the reader these sections for precise statements. These results
provide estimates of the drop in terms of prescribed values. For example (Corollary \ref{sessile-co}) 

\begin{quote} {\it  Let $\Pi$ be a spacelike plane in $\l^3$. Consider $\beta, R>0$. Then for any 
stationary sessile liquid drop $X$ resting on $\Pi$ such that $\beta$ is  the angle of contact  and $R$ is the radius of the disc
of $\Pi$ that wets $X$, the height $q$ of $X$ satisfies:
$$q<R\frac{\cosh\beta-1}{\sinh\beta},$$}
\end{quote}
We point up also an existence result in terms of the volume (Theorems \ref{sessile-vo} and \ref{pendent-vo}) 

\begin{quote} {\it  For  each  constant of capillarity $\kappa$, $\beta\in\r$ and 
a  positive real number ${\cal V}$, there exists a unique sessile liquid drop and a unique 
pendent drop resting on a spacelike plane $\Pi$ enclosing a liquid of volume ${\cal V}$ and that makes a constant angle $\beta$ along the 
liquid-air-solid interface.}
\end{quote}

We want to point out although much of our techniques are similar than in Euclidean ambient, 
differences appear in the Lorentzian setting. The main fact is that the spacelike condition 
 is a strong geometric restriction for the possible configurations. It is worthwhile to bring out some of them:

\begin{enumerate}
\item Liquid drops in $\l^3$ can  extend to be graphs in the whole plane.
This means that the associated Euler equation can solved be entire solutions. Our drops correspond with physical situations of 
wetting, but dewetting is prohibited. 

\item Sessile liquid drops in $\l^3$ have not the phenomenon of meniscus. Moreover and by fixing the 
constant of capillarity, 
the ambient space $\l^3$ can be foliated by sessile drops.

\item Pendent liquid drops in $\l^3$ do not present vertical points. In particular, for any given volume, there exists
a  pendent drop enclosing that volume and that it is physically realizable.
\end{enumerate}

There exists an extensive literature relating to liquid drops in Euclidean space. This is due to its interest both in physic and 
chemistry for any interfacial phenomena in the theory of  colloids and new materials.
From the mathematical viewpoint, we refer to  
the book of R. Finn \cite{fi1}, which contains abundant bibliography.
Finally, we remark that much of the results obtained here can be straightforward generalize for 
spacelike hypersurfaces in $\l^{n+1}$.

%%%%%%%%%%%%%%%%%%%%%%%%%%%%%%%%%%%%%%%%%
\section{Preliminaries and formulation of the variational problem}\label{sec2}
%%%%%%%%%%%%%%%%%%%%%%%%%%%%%%%%%%%%%%%%%%%%%%

In this section we present the setting for the presence of our stationary 
liquid drops, as well as, we shall precise 
the mathematical formulation of the physical situation. 
Let $\l^3$ denote the 3-dimensional Lorentz-Minkowski space, that is, the real vector space $\r^3$ endowed with the 
Lorentzian metric $\langle,\rangle=dx_1^2+dx_2^2-dx_3^2$, where $x=(x_1,x_2,x_3)$ are the canonical 
coordinates in $\l^3$. An  immersion $x: M\rightarrow \l^3$ of a smooth surface 
$M$  is called {\it spacelike} if the induced metric on $M$ is positive definite. 
Observe that $\vec{a}=(0,0,1)$ is a unit timelike vector field globally defined on $\l^3$, which determines 
a time-orientation on the space $\l^3$. This allows us to choose a unique unit normal vector field $N$ on $M$ which is in the 
same time-orientation as $\vec{a}$, and hence that $M$ is oriented by $N$. We will refer to 
$N$ as the future-directed Gauss map of $M$. In this 
article  all spacelike surfaces  will be oriented according to this choice of $N$.

The spacelike condition  imposes topological restrictions to the immersion $x$. For example, 
 there are not closed spacelike surfaces and then, any 
compact spacelike surface   has non-empty 
boundary. If $\Gamma$ is  a closed curve in $\l^{3}$ and  $x: M\rightarrow\l^{3}$ is a 
spacelike immersion of a compact surface, we say that the boundary of $M$ is  $\Gamma$ if the restriction  $x:\partial  M
\rightarrow \Gamma$ is a diffeomorphism. 
For spacelike surfaces, the projection $\pi:\l^3\rightarrow \Pi=\{x_3=0\}$, $\pi(x_1,x_2,x_3)=(x_1,x_2,0)$ is  a local diffeomorphism 
between $\mbox{int}(M)$ and $\pi(\mbox{int}(M))$. Thus, $\pi$ is 
an open map and $\pi(\mbox{int}(M))$ is a domain in $\Pi$. The compactness of $M$ implies that $\pi:M\rightarrow\overline{\Omega}$ is a covering map. Thus, we have

\begin{proposition}\label{p1}
Let $x:M\rightarrow\l^3$ be  a compact spacelike surface whose boundary 
$\Gamma$ is a graph over  an open region $\Omega\subset\{x_3=0\}$. Then $x(M)$ is  a graph
over $\Omega$. 
\end{proposition}

As conclusion, the boundness of our liquid drops means that the liquid-air interface ${\cal S}$ is a graph 
on the support surface. We suppose that the boundary of a spacelike compact surface is included in a plane $\Pi$. This 
plane  must be of  spacelike-type. We point out also that although the boundary $\partial M$ is possibly non-connected,
 the causal character on $M$
implies the existence of a component  of 
$x(\partial M)$, named $\Gamma_{0}$, such that $\pi(\mbox{int}( M))$ is contained in the bounded domain 
determined by $\Gamma_{0}$ in $\Pi$. Therefore,   $x(M)$ defines an "interior" domain, that is, there exists a bounded region $\Omega\subset \Pi$
such that $x(M)\cup \Omega$ determines in $\r^3$ a bounded domain $B$, called the "interior" of $M$. 

For spacelike immersions, the notions of 
the first and  second fundamental form are defined in the same way as  in Euclidean space.
In classical notation, the first and the second fundamental form of $x$ are
$${\rm I}=\sum_{i j}g_{ij} dx_i\ dx_j,\hspace*{1cm}{\rm II}=\sum_{i j} h_{ij} dx_i\ dx_j,$$
where $g_{ij}=\langle\partial_i x,\partial_j x\rangle$ is the induced metric on $M$ by $x$ and 
$h_{ij}=\langle \partial_i N,\partial_j x\rangle$.
Then the  mean curvature $H$ of $x$ is given then by 
$$2H=\frac{h_{22} g_{11}-2h_{12} g_{12}+h_{11}g_{22}}{\mbox{det}(g_{ij})}.$$
Assume that   $M$ is the graph of a smooth function $u=u(x_1,x_2)$ 
defined over a domain $\Omega$.
 The spacelike condition implies $|\nabla u |<1$, where $\nabla$ is the gradient operator in $\r^2$ and 
the Gauss map is 
$$N=\frac{(\nabla u,1)}{\sqrt{1-|\nabla u|^2}}.$$
According this orientation, the mean curvature $H$ at the point 
$(x,u(x))$,  $x\in\Omega$,  satisfies the equation
$$(1-|\nabla u|^2)\Delta u-\sum u_i u_j u_{ij}=2H(1-|\nabla u|^2)^{3/2}.$$
This equation can alternatively be written in divergence form
\begin{equation}\label{media}
\mbox{div}(Tu)=2 H,\hspace*{1cm}Tu=\frac{\nabla u }{\sqrt{1-|\nabla u |^2}}.
\end{equation}

We present   now the notion of stationary surface in $\l^3$. The {\it support surface} $\Sigma$ is defined as 
 an embedded connected spacelike surface in $\l^3$ that  divides the space $\l^3$ into two connected components. 
Let us orient $\Sigma$ by the future-directed unit timelike vector field $N_{\Sigma}$ and consider 
$\l^3_+$ the component of $\l^3\setminus\Sigma$ towards $N_{\Sigma}$ is pointing. 
Let $M$ be a connected compact surface with boundary $\partial M$ and 
$x: M\rightarrow\l^3$ a  spacelike immersion, smooth even at $\partial\Sigma$ such that $x(\mbox{int}(M))\subset 
\l^3_+$ and $x(\partial M)\subset \Sigma$. 
A {\it variation} of $x$ is a differentiable map $X:(-\epsilon,\epsilon)\times M\rightarrow\l^3$ such that 
$X_t:M\rightarrow\l^3$, $t\in(-\epsilon,\epsilon)$, defined by $X_t(p)=X(t,p)$, $p\in M$, is an immersion
and $X_0=x$. The variation is called {\it admissible} if   $X_t(\mbox{int}(M))\subset\l^3_+$ and 
$X_t(\partial M)\subset \Sigma$ for all $t$.
The {\it area function} $A:(-\epsilon,\epsilon)\rightarrow\r$ is defined by 
$$A(t)=\int_M d A_t$$
where $d A_t$ is the area element of $M$ in the metric induced by $X_t$; the area function $S:(-\epsilon,\epsilon)
\rightarrow\r$ of the wetted surface on $\Sigma$ is defined by 
$$S(t)=\int_{\Omega_t} d\Sigma,$$
that is, the area of $\Omega_t\subset\Sigma$, the region in $\Sigma$ bounded by $X_t(\partial M)$. Finally, the {\it volume function} 
$V:(-\epsilon,\epsilon)\rightarrow\r$ is defined by 
$$V(t)=\int_{[0,t]\times M}X^*\ dV,$$
where $d V$ is the canonical volume element of $\l^{3}$. The number $V(t)$ represents the volume enclosed between the 
surface $X$ and $X_t$. The variation $X$ is said to be {\it volume-preserving} if $V(t)=V(0)$ for all $t$ and 
the {\it variational vector field }
of $X$ is defined on $M$ by 
$$\xi(p)=\frac{\partial X}{\partial t}(p){\biggl|}_{t=0}.$$
Moreover, we assume the existence of a potential energy $Y=Y(p)$, $p\in\l^3$.  The resultant variation
energy is 
$$Y(t)=  \int_M Y\ dA_t.$$

The energy function $E:(-\epsilon,\epsilon)\rightarrow\r$ is defined by 
$$E(t)=A(t)-\cosh\beta\ S(t)+Y(t),$$
where $\beta\in\r$ is an arbitrary real constant. We say that the immersion $x$ is {\it stationary} if $E'(0)=0$ for any volume preserving admissible variation of $x$. For an arbitrary variation, it can be shown that 
\begin{eqnarray*}
A'(0)&=&-2\int_M H\langle N,\xi\rangle\ dM-\int_{\partial M}\langle\nu,\xi\rangle\ ds\\
S'(0)&=&-\int_M\langle\nu_{\Sigma},\xi\rangle\ ds\\
Y'(0)&=&\int_M Y\langle N,\xi \rangle\ dM,
\end{eqnarray*}
where $\nu$ and $\nu_{\Sigma}$ are  the inward-pointing unitary conormal to $M$ and $\Omega$ along $\partial M$
respectively. Moreover, the first variation formula of the volume is given by (cf. \cite{bo1,bo2})
$$V'(0)=-\int_M\langle N,\xi\rangle dA.$$
 Hence we obtain the first variation formula for the energy of the physical system:
$$E'(0)=\int_M\left(-2H +Y+\lambda\right)\langle N,\xi\rangle \ dM+\int_{\partial M} \langle 
\xi,\nu_{\Sigma}\rangle\left(\cosh\beta+\langle N,N_{\Sigma}\rangle\right)\ ds$$
Thus, we have

\begin{proposition} Let $\Sigma$ be a support surface in $\l^3$ and let $M$ be  
a compact surface. Let us consider $x:M\rightarrow\l^3$  a smooth 
spacelike immersion such that $x(int(M))\subset \l^3_+$ and  $x(\partial M)\subset\Sigma$. Then $x$ is stationary if 
and only if 
\begin{enumerate} 
\item The mean curvature $H$ of $x$ satisfies the relation 
$$2H(p)=Y(p)+\lambda,\hspace*{1cm} p\in M,$$
where $Y$ is a potential energy and $\lambda$ is a Lagrange parameter determined by an eventual volume 
constraint;
\item The surface ${\cal S}=x(M)$ meets the support surface $\Sigma$ in a constant  hyperbolic angle $\beta$,
that is, $\cosh\beta=-\langle N,N_{\Sigma}\rangle$ along $\partial M$.
\end{enumerate}
\end{proposition}

In this article, our interest will center on the case for which: 
\begin{enumerate}

\item The support surface is a spacelike hyperplane $\Pi$. After an isometry, we will assume that  $\Pi$ 
is parallel to the plane $\{x_3=0\}$ and 
\item The vector field $Y(p)$ is a vertically directed gravitational potential towards $\Pi$, that is, 
$Y(p)=\kappa\ x_3(p)+\lambda$, for constants  $\kappa$ and $\lambda$:
\end{enumerate}

We know from Proposition \ref{p1} that a compact drop resting in spacelike plane is the graph of a function $u$. 
In such case, 
$$-\langle N,N_{\Pi}\rangle=\frac{1}{\sqrt{1-|\nabla u|^2}}=\cosh\beta\hspace*{1cm}\mbox{along }\partial M.$$
Moreover, the constancy of the hyperbolic angle along $\partial M$ means that the Euclidean angle is also constant 
along this curve since, $|\nabla u|$ is constant along $\partial M$ and 
$$\langle N^E,\vec{a}{\rangle}_{E}=\frac{1}{\sqrt{1+\tanh^2\beta}}\hspace*{1cm}\mbox{ along }\partial M,$$
where $N^E$ and $\langle,\rangle_E$ denote, respectively, the Euclidean unit normal of $M$ and 
the Euclidean metric of $\r^3$.

Since we shall study stationary liquid drops, throughout this work we shall omit the word 
stationary and it is implied that a liquid drop is a {\it stationary } liquid drop.

%%%%%%%%%%%%%%%%%%%%%%%%%%%%%%%%%
\section{Symmetry under gravitational fields}\label{cap-wente}
%%%%%%%%%%%%%%%%%%%%%%%%%%%%%%%%%

In this section we prove  that the equilibrium shape of a drop of liquid 
in $\l^3$ resting over a spacelike plane  in a uniform gravitational field 
is rotational symmetric with respect to a straight-line orthogonal to 
the support surface. Exactly we show

\begin{theorem}\label{t1} Let $M$ be a spacelike embedded compact connected surface in $\l^3$ 
whose boundary $\partial M$ is contained in a plane $\Pi$. Assume that $ M$ lies 
in one side of $\Pi$ and the two following assumptions hold:
\begin{enumerate} 
\item The mean curvature $H$ of 
$ M$ depends only on the distance to $\Pi$.
\item The hyperbolic angle of contact between $ M$ and $\Pi$ is constant along $\partial M$. 
\end{enumerate}
Then $ M$ is rotational symmetric with respect to  a straight-line orthogonal to $\Pi$. Moreover, 
$M$  is   a topological disc.
\end{theorem}

This result is analogous to it happens in Euclidean ambient and which  was proved by Wente \cite{we}. 
It turns out that the method of proof used there,  called the  Alexandrov reflection technique, 
may be adapted to the present situation. Such technique  was firstly  used  to prove that a closed embedded constant mean curvature surface in $\r^3$ must be a round sphere \cite{al}. See also the remarkable reference \cite{se} in the
  context of the theory of partial differential equations. 
 The proof  idea is to use the very surface $M$ as comparison surface with itself and to 
apply the Hopf maximum principle for elliptic equations.

We consider $u^i$, $i=1,2$, two functions satisfying
$$\mbox{div}(Tu^i)= 2 H_i(x),\hspace*{1cm}\ |\nabla u^i|<1$$
in a domain $\Omega$, with $H_1(x)\leq H_2(x)$. The  operator $\mbox{div}(Tu)$  may be written in the form
$$\mbox{div}(Tu)=\frac{(1-u_2^2)u_{11}+2 u_1 u_2 u_{12}+(1-u_1^2)u_{22}}
{W^3},\hspace*{1cm}W=\sqrt{1-|\nabla u|^2}$$
where  the subscript $i$  indicates the differentiation with respect to the variable $x_i$. We write
$\mbox{div}(Tu)=\sum a_{ij}(x, u,\nabla u)u_{ij}$ with $p=(p_1,p_2)$, $p_i=u_i$, and where
$$\sum a_{ij}(x,u,p)\xi_i\xi_j=\frac{(1-|p|^2)|\xi|^2+\langle\xi,p\rangle^2}{W^3}.$$
 Then it holds
$$0<\lambda(x,u,p)|\xi|^2\leq \sum a_{ij}(x,u,p)\xi_i\xi_j\leq\Lambda(x,u,p)|\xi|^2,$$
$$\lambda(x,u,p)=\frac{1}{W}\hspace*{1cm}\Lambda(x,u,p)=\frac{1}{W^3}.$$
Then the operator is elliptic for $|p|<1$ and uniformly elliptic for compact domains.
Let
\begin{equation}\label{media2} \phi(x,p,r)=\mbox{div}(Tu)=2 H(x)\end{equation}
where $r=(r_{ij})$, $r_{ij}=u_{ij}$. Then 
$\phi$ is a smooth function defined in $\Omega\times D\times\r^4$ given explicitly by  
$$\phi(x,p,r)=\frac{1}{\sqrt{1-|p|^2}}\sum\left(\delta_{ij}+\frac{p_i p_j}{1-|p|^2}\right) r_{ij},$$
where $D$ is the unit open disc of $\r^2$. 
For each  $u=u^i$, $i=1,2$, we will use the notation $p^i, r^i$ and $H_i$ for each $i$ . Since  $H_1\leq H_2$, a
 standard argument  using the chain rule shows then 
\begin{eqnarray*}
0&\leq & \phi(x,p^2,r^2)-\phi(x,p^1,r^1)\\
&=& \sum\int_0^1\frac{\partial \phi}{\partial r_{ij}}(\theta(t))\ dt\ w_{ij}+
\sum\int_0^1\frac{\partial \phi}{\partial p_j}(\theta(t))\ dt\ w_j:=Lw,
\end{eqnarray*}
where $w=u^1-u^2$, $w_i=\partial w/\partial x_i$, $w_{ij}=\partial^2 w/\partial x_i\partial x_j$ 
and $\theta=\theta(t)=(x,tp^2+(1-t)p^1,t r^2+(1-t)r^1)$. 
The right  hand side of the above equation defines an elliptic operator $L$ because 
$$ |\xi|^2\leq \int_0^1 \frac{1}{W(\theta(t))}\ dt|\xi|^2\leq Lw\leq \max\left\{\frac{1}{W_1^3},\frac{1}{W_2^3}\right\}|\xi|^2,$$
and $W_i=\sqrt{1-|\nabla u^i|^2}$. Since the coefficients $a_{ij}$ are locally bounded, 
$L$ is locally  uniformly elliptic and we are in position to apply the Hopf's maximum principle to the difference function $w$ (\cite{ho}; see also \cite[Ch. 3]{gt}). Consequently, we have proved the following result.

\begin{theorem}[The touching principle] \label{touching}
Let $u,v$ be two  smooth solutions to the 
same prescribed mean curvature equation (\ref{media2}) on a domain $\Omega\subset\r^2$.
 Suppose that $u\leq v$ on $\Omega$ and 
$u(x_0)=v(r_0)$, $x_0\in\Omega$. Then $u(x)=v(x)$ on $\Omega$. 
The same holds if $p\in\partial\Omega$ with the extra hypothesis that 
$\partial u/\partial\nu=\partial v/\partial \nu$ at $x_0$, where $\nu$ is the outward unit normal to $\partial\Omega$.
\end{theorem}

\begin{proof}[of Theorem \ref{t1}]  After an ambient isometry, we can suppose that 
$\Pi\equiv\{x_3=0\}$ and $H(x)=H(x_3(x))$ for any $x\in M$. Without loss of generality, we assume that $ M$ lies over the plane $\Pi$. The proof follows the same steps of that of Wente \cite[Th. 1.1]{we} and for expository reasons, we will describe it briefly.

Let $\Omega$ be the bounded region  in $\Pi$ bounded by $\partial M$ such that $M\cup \Omega$ is a closed embedded surface.  Let $A$ and  $B$ denote, respectively, 
the nonbounded and the interior domain determined by $M$ in $\r^3$. 
 Consider a vertical hyperplane $P$ disjoint from $ M$ (so $P\subset A$) and move $P$ 
parallel to itself (say, to the right) until it touches $ M$ at a first point $q$. Now, when moving $P$ a little more to the right from $q$, to a plane $P(t)$, the (closed) part of $ M$ on the left of $P(t)$, which we denote by $ M(t)^-$, is a graph (with 
respect to the horizontal) over a domain in $P(t)$ and no point of $ M(t)^-$ has a horizontal tangent plane.

Let $ M(t)^+$ be the symmetry of $ M(t)^-$ through $P(t)$ that it is contained in $B$. Recall that the symmetry 
with respect to  a vertical plane is an isometry of $\l^3$, and so, the mean curvature remains invariant by the 
symmetry. Because 
the mean curvature of $M$ depends only on the height with respect to $\Pi$, the mean curvature is the same 
for all points at the same height. 
We continue now moving $P(t)$ to the right, and consider the 
first parallel plane $P(\tau)$ where one of the following conditions fails to hold:
\begin{enumerate}
\item $int( M(\tau)^+)\subset int(B)$.
\item $ M(\tau)^-$ is a graph over a part of $P(\tau)$ and no point of $ M(\tau)^-$ has a horizontal tangent plane.
\end{enumerate}
If 1) fails first, we have that $M$ and $M(\tau)^+$ touch at some interior point $p$, or at a 
boundary point $p$, $p\in\partial M\cap\partial M(\tau)^+$ where the constancy of the hyperbolic  angle along $\partial M$ 
implies that the tangent planes of $M$ and $M(\tau)^+$ agree at $p$. Then
 one applies the touching  principle to $ M$ and $ M(\tau)^+$ at the point where 
they touch to conclude that $P(\tau)$ is a plane of symmetry of $ M$. 

If 2) fails first, then we have that there exists a point 
$p$ where the tangent plane of $ M(\tau)^-$ becomes horizontal is on $\partial( M(\tau)^-)\subset P(\tau)$ or
$p\in\partial M\cap P(\tau)$. In the former possibility  one can apply the boundary touching  principle to $ M(\tau)^+$ and the part of $ M$ to the right to $P(\tau)$ to conclude that $P(\tau)$ is a plane of symmetry of $ M$; in the second one,  the corresponding tangent planes of $M$ and 
$M(\tau)^+$ are identical because the hyperbolic angle with the $\vec{a}$ direction  is the same at $p$. Then one applies 
the maximum principle at a corner point (see details in \cite{se,we}).

Thus, for each vertical plane $P$, some parallel 
translate of $P$ is a plane of symmetry of $ M$, and $ M$ is a surface of revolution. This concludes the proof of Theorem.

 \hfill{$q.e.d$}
\end{proof}
Finally  we comment other situations where is is possible to apply the 
Alexandrov reflection method. The physical problem we will consider is that a drop of liquid is trapped between two 
homogeneous parallel spacelike planes $\Pi_1$ and $\Pi_2$, that is, a bounded liquid bridge $M$. In such case, 
the term $S$ in the energy functional $E$ is the area of the domains that the drop wets in each one of the planes.
 Again, in a state of   equilibrium, 
the angle between the normal to the liquid bridge and the normal to $\Pi_i$ along their line of contact is 
constant (and possibly different in each plane $\Pi_i$). The Alexandrov reflection method yields again the following  

\begin{theorem} \label{t2} Let $\Pi_1$ and $\Pi_2$ be two parallel spacelike planes
in Minkowski space $\l^3$. 
Consider $M$ a spacelike embedded compact  surface in $\l^3$ included in the 
slab determined by $\Pi_1\cup\Pi_2$ and 
whose boundary $\partial M$ intersects both $\Pi_1$ and $\Pi_2$. 
 Assume that the mean curvature of 
$M$ depends only on the distance to $\Pi_i$ and  the hyperbolic angle of contact between $M$ and $\Pi_i$ 
 is constant along $\partial M$ in each one of the two planes. 
Then $M$ is   rotational symmetric with respect to  a straight-line orthogonal to $\Pi_i$.
\end{theorem}

\begin{remark} {\rm Other interested support  surface occurs when $\Sigma$ is  a hyperbolic plane. One can believe that the only stationary 
liquid drops resting on a hyperbolic plane are surfaces of revolution. This is true in the 
situation of no gravity, that is, when the mean curvature of the  liquid-air interface is constant. In such case, we obtain 
that  the surface must be an umbilical disc \cite{ap}. We do not know if the same remains true 
under the effect of a gravitational vector field.}
\end{remark}

\begin{remark} {\rm   The study of 
constant mean curvature surfaces in Euclidean space that makes a constant angle with a 
prescribed support surface $\Sigma$ is a focus of interest in differential geometry. 
When $M$ is a topological disc,
there are results that assure that 
$M$ is  a spherical cap  or a planar disc, for example, provided  $\Sigma$ is a sphere \cite{ni}
or  a plane \cite{lo}. The arguments use the holomorphicity of the Hopf differential. Generalizations exist
for other ambient spaces \cite{ap,so}. See also \cite{ko,rv} when we impose conditions on stability.}
\end{remark}

%%%%%%%%%%%%%%%%%%%%%%%%%%%%%%%%%
\section{Existence and uniqueness of solutions}\label{cap-exi}
%%%%%%%%%%%%%%%%%%%%%%%%%%%%%%%%%%%%

From Theorem \ref{t1}, we know  that a stationary liquid drop resting in a spacelike plane $\Pi$ and under the effect of 
a gravitational field $Y$ as linear function of the height with respect to $\Pi$ must be a surface of revolution. Moreover, Proposition \ref{p1}  assures that the 
drop must be a graph over a domain of $\Pi$. Thus the corresponding Euler equation that describes the shape of a stationary
liquid drop becomes an ordinary differential equation. In this section, we shall study the problem of existence and 
uniqueness of such equation.

We may assume without loss of generality  that the support plane $\Pi$ is parallel to the plane $\{x_3=0\}$ and that
the potential energy is 
$Y=\kappa\ x_3+\lambda$, $\kappa,\lambda\in\r$. Let $M$ be a rotational symmetric spacelike graph supported 
in $\Pi$ whose  mean curvature is  $Y$ and 
that makes a hyperbolic angle $\beta$ with $\Pi$  along 
$\partial M$. After a horizontal translation, we assume that the axis of revolution is the $x_3$-axis. 
Then  $M$ is the graph of a function $u: [0,R)\rightarrow\r$ and it can be represented as $M=\{r \cos\theta,r\sin\theta,u(r));r\in [0,R),\theta\in\r\}$. 
 Equation (\ref{media}) converts into 
\begin{equation}\label{rotational1}
\frac{1}{r}\frac{d}{dr}\left(\frac{r u'(r)}{\sqrt{1-u'(r)^2}}\right)=\kappa\  u(r)+\lambda,\hspace*{1cm} 0\leq r<R,
\end{equation}
with boundary conditions
\begin{equation}\label{rotational11}
u'(0^+)=0,\hspace*{1cm}u'(R^-)=\tanh\beta.\end{equation}
If $\kappa=0$, the solutions correspond to a constant mean curvature surfaces, with $H=\lambda/2$. 
A direct integration of (\ref{rotational1})  leads to 
$u(r)=\pm\sqrt{r^2+\frac{4}{\lambda^2}}+c$, if $\lambda\not=0$ and $u(r)=c$, if $\lambda=0$. In the 
first case, $u$ describes a hyperbolic plane of mean curvature $\lambda/2$; in the second one, we obtain 
a horizontal plane parallel to $\Pi$. As conclusion, we have

\begin{proposition} Let $\Pi$ be a support spacelike plane and $\beta$ a real number. If we assume no gravity in the ambient 
space $\l^3$, then there exists a unique stationary liquid drop resting on $\Pi$ and that makes a constant angle $\beta $ of contact 
along the liquid-solid-air interface. Moreover, one can prescribe the area of the wetted region on $\Pi$ by the drop.
\end{proposition}

\begin{proof} Assume $\Pi=\{x_3=c\}$. If $\beta =0$, it suffices to take $u(r)=c$. Let $\beta\not=0$ and  a positive number $R$. It suffices to define 
$$u(r)=\frac{\beta}{|\beta|}\sqrt{r^2+\frac{R^2}{\sinh^2\beta}}+c-R\coth\beta,\ r\in [0,R).$$

\hfill{$q.e.d$}
\end{proof}
After this result, we will suppose in the rest of this work that  $\kappa\not=0$. 
The transformation $u\rightarrow u-\frac{\lambda}{\kappa}$  changes (\ref{rotational1}) into 
\begin{equation}\label{rotational2}
\frac{d}{dr}\left(\frac{r u'(r)}{\sqrt{1-u'(r)^2}}\right)=\kappa\ r u(r),\hspace*{1cm}0\leq r<R.
\end{equation}
Let $u(r)\rightarrow \sqrt{|\kappa|}u(r/\sqrt{|\kappa|})$. Equation (\ref{rotational2})  writes now as
\begin{equation}\label{unique}
\frac{d}{dr}\left(\frac{r u'(r)}{\sqrt{1-u'(r)^2}}\right)=\epsilon\ r u(r),\ 0\leq r<\sqrt{|\kappa|R},
\end{equation}
with $\epsilon=1 $ (resp. $-1$) if $\kappa$ is positive (resp. negative).

We attack the problem of existence and uniqueness of stationary liquid drops by studying  the initial value problem
\begin{eqnarray}
& &\frac{d}{dr}\left(\frac{r u'(r)}{\sqrt{1-u'(r)^2}}\right)=\epsilon\ r u(r),\ 0\leq r< b,\label{ivp-1}\\
& &u(0^+)=u_0\hspace*{1cm}u'(0^+)=0\label{ivp-2}
\end{eqnarray}
We denote $u=u(r;u_0)$ the dependence on the initial boundary condition $u(0^+)=u_0$. 
Because of the singularity of (\ref{ivp-1}) at $r=0$, standard existence theorems for differential equations cannot 
be used to show even local existence or uniqueness of solutions of (\ref{ivp-1})-(\ref{ivp-2}).
However the uniqueness can be established as follows.

\begin{theorem}[Uniqueness]\label{t-uni} Let $u(r;u_0)$ and $u(r;u_1)$ be solutions of (\ref{ivp-1})-(\ref{ivp-2}).
Then there exists a constant $M=M(b)$ such that 
\begin{equation}\label{eq-m}
|u(r;u_0)-u(r;u_1)|\leq M |u_0-u_1|
\end{equation}
for $0\leq r < b$. In particular, the initial values problem (\ref{ivp-1})-(\ref{ivp-2})
 has at most one solution.
\end{theorem}

\begin{proof} Define $\psi=\psi(r)$ by $u'(r)=\tanh\psi$. Then
Equation (\ref{ivp-1}) writes as 
$(r\sinh\psi)'=\epsilon\  r u(r)$, with $\sinh\psi=u'/\sqrt{1-u'^2}$. Integrating from $0$ to $r$, we obtain
\begin{equation}\label{eq-sinh}\sinh\psi(r)=\frac{1}{r}\int_0^r \epsilon\  t u(t)\ dt.
\end{equation}
 Because the function $x/\sqrt{1+x^2}$ has Lipschitz constant $1$, we have
\begin{eqnarray*}|u(r;u_0)-u(r;u_1)|&\leq &|u_0-u_1|+
\int_0^r|\tanh(\psi_1)-\tanh(\psi_2)|\ dt\\
&\leq &|u_0-u_1|+
\int_0^r|\sinh(\psi_1)-\sinh(\psi_2)|\ dt\\
&\leq& |u_0-u_1|+
\int_0^r\frac{1}{t}\left(\int_0^t s| u(s;u_0)-u(s;u_1)|ds\right)\ dt,
\end{eqnarray*}
where in the last inequality we have used (\ref{eq-sinh}) and $|\epsilon|=1$. By interchanging the order of integration, we obtain
$$|u(r;u_0)-u(r;u_1)|\leq |u_0-u_1|+\int_0^r s \log\frac{r}{s} |u(s;u_0)-u(s;u_1)|\ ds.$$
Since $0\leq r<b$, and by the behaviour of the function $ x\log(b/x)$, we have 
$$|u(r;u_0)-u(r;u_1)|\leq |u_0-u_1|+\frac{b}{e}\int_0^r  |u(s;u_0)-u(s;u_1)|\ ds.$$
By applying  the Gronwall's lemma, we obtain 
$$|u(r;u_0)-u(r;u_1)|\leq |u_0-u_1| e^{br/e}\leq |u_0-u_1|e^{b^2/e}.$$
By setting $M=e^{b^2/e}$, we conclude the proof.

\hfill{$q.e.d$}
\end{proof}
We study now the existence of (\ref{ivp-1})-(\ref{ivp-2}). If $u_0=0$, the solution  is $u=0$. We assume then $u_0\not=0$.
We can adapt to our situation the method of majorants to 
obtain local existence (if 
  $\epsilon=1$, we could  consider  the method of successive approximations as the Euclidean case developed 
in \cite{jp}).  The advantage of this proof is that we obtain in addition to the continuity on the parameters.
Since $u(r;u_0)=-u(r;-u_0)$,  it suffices to consider $u_0>0$. 
 
\begin{theorem}[Existence] \label{t-exi}
Given $u_0>0$,  there exists a (unique) solution $u$ of (\ref{ivp-1})-(\ref{ivp-2}). The solution $u=u(r;u_0)$ 
depends analytically on the parameter $u_0$ and the maximal interval of definition of $u$ is $[0,\infty)$.
\end{theorem}

\begin{proof}We follow the same ideas than in Euclidean space \cite{we} and  we  only describe sketch the idea.
Put $v=u'/\sqrt{1-u'^2}$. Equation (\ref{ivp-1}) becomes
\begin{equation}\label{eq-v}
(rv)'=\epsilon ru\hspace*{.5cm}\mbox{or}\hspace*{.5cm} v'+\frac{v}{r}=\epsilon u.
\end{equation} Also (\ref{eq-sinh}) is now 
\begin{equation}\label{ex1}
v(r)=\frac{\epsilon}{r}\int_0^r t u(t)\ dt.
\end{equation}
 By differentiation of (\ref{eq-v}), we obtain 
\begin{equation}\label{uve}
(rv')'-v\left(\frac{\epsilon r}{\sqrt{1+v^2}}+\frac{1}{r}\right)=0.
\end{equation}
Moreover (\ref{ex1}) yields $\lim_{r\rightarrow 0} (v/r)=\epsilon u_0/2$ (also cf. (\ref{B-2}) and (\ref{P-2})).
Thus $\lim_{r\rightarrow 0}v'(r)=\epsilon u_0/2$.
Then (\ref{ivp-1})-(\ref{ivp-2}) is equivalent to find a solution $v$ of (\ref{uve}) with initial values
$$v(0)=0,\hspace*{1cm}v'(0)=\epsilon u_0/2.$$
 Define the differential operators
$$L(v)=(rv')'-\frac{v}{r},\hspace*{1cm}M(v)=\frac{\epsilon r v}{W},\ W=\sqrt{1+v^2}.$$
We write $v(r)=\sum_{n=1} a_n r^n$ in a potential series, with $a_1=\epsilon u_0/2$. The radius of convergence for $M(v)$ is $1$

Define 
$$L_1(w)=w'-\frac{\epsilon u_0}{2}\hspace*{1cm}M_1(w)=Cr\left(\frac{w/\rho}{1+w/\rho}\right),$$
for an appropriate constant $C$ depending on the convergence of $M$. 
For a function $w=\sum_{n=1}b_n r^n$, with $w'(0)=|a_1|$, 
we consider the auxiliary problem $L_1(w)=M_1(w)$. One proves then that 
the series for a solution $w$ of this problem will majorize the series for $v$ with $L(v)=M(v)$.
Clearly  the initial value problem 
$$L_1(w)=w'-\frac{\epsilon  u_0}{2}=C r\left(\frac{w/\rho}{1+w/\rho}\right)=M_1(w)$$
$$w(0)=0$$
can be solved by a convergent power series $w$ that majorizes the series for  $v$ 
in the circle of convergence. This implies that $v$ is also convergent. Since the solution $v$ writes as an
series of potential, the solution $u$ of (\ref{ivp-1})-(\ref{ivp-2})  is analytic in the parameter $(r;u_0)$.

For the study of the maximal interval for a solution $u$,  we transform the initial value problem (\ref{ivp-1})-(\ref{ivp-2}) into a pair of integral equations 
\begin{eqnarray}
& &u(r)=u_0+\int_0^r\frac{v(t)}{\sqrt{1+v(t)^2}}dt\label{eq2-a}\\
& &v(r)=\epsilon \frac{1}{r}\int_0^r   t u(t)\ dt.\label{eq2-b}
\end{eqnarray}
with $u(0)=u_0$, $v(0)=0$. It is straightforward that if $u$ and $v$ are continuous functions that satisfy such integral equations, then 
$u'$ is continuous and $u$ is a solution of (\ref{ivp-1})-(\ref{ivp-2}). The integrand in (\ref{eq2-a}) is continuous for any $r$. This allows to extend the solution $u(r;u_0)$ until $\infty$.

\hfill{$q.e.d.$}
\end{proof}

As consequence of Theorems \ref{t-uni} and \ref{t-exi}, we conclude 

\begin{corollary} Consider in Lorentz-Minkowsk space $\l^3$ a spacelike plane $\Pi$ and a 
gravity vector field orthogonal to $\Pi$. Then any stationary liquid drop of $\l^3$ supported on $\Pi$ 
 can extend to be a graph defined in the whole $(x_1,x_2)$-plane. 
\end{corollary}

Although in this section we have studied the problem of existence of (\ref{ivp-1})-(\ref{ivp-2}), this  
allows to solve the problem of 
existence of the boundary problem (\ref{rotational1})-(\ref{rotational11}). This will be proved in 
Corollaries \ref{sessile-ex} and \ref{pendent-ex}).

We end this section with a  property on the dependence of a solution of the 
stationary liquid drop equation (\ref{rotational1}) with respect to the capillarity constant $\kappa$.

\begin{theorem}[Monotony with respect to $\kappa$] \label{sessile-mono} Let $\kappa_1,$ and $\kappa_2$ be two positive constants of capillarity. Denote $u_i=u_i(r)$, $i=1,2$,  two solutions of  (\ref{rotational1})-(\ref{rotational11}) for $\kappa=\kappa_i$ and where 
$u_i(0)>0$. If $\kappa_1<\kappa_2$, then 
\begin{enumerate}
\item $u_1(r)>u_2(r)$ for $0\leq r\leq R$.
\item  $u_1'(r)>u_2'(r)$ for $0<r<R$. 
\end{enumerate}
\end{theorem}

\begin{proof} After a vertical translation, we assume $\lambda=0$ in (\ref{rotational1}).
Set $v_i=u_i'/\sqrt{1-u_i'^2}$. Then  $(rv_i)'=\kappa_i r u_i$. For 
each $0\leq r_0<r<R$, we integrate between $r=r_0$ and $r=R$:
$$r v_i(r)-r_0 v_i(r_0)=\kappa_i\int_{r_0}^r t u_i(t)\ dt.$$
Then
\begin{equation}\label{mono1}
r\Bigl(v_2(r)-v_1(r)\Bigr)=r_0\Bigl(v_2(r_0)-v_1(r_0)\Bigr)+\int_{r_0}^r t(\kappa_2 u_2(t)-\kappa_1 u_1(t)) \ dt.\end{equation}
For $r=R$ and because $u_1'(R)=u_2'(R)$, 
\begin{equation}\label{mono2}
r_0\Bigl(v_1(r_0)-v_2(r_0)\Bigr)=\int_{r_0}^R t\Bigl(\kappa_2 u_2(t)-\kappa_1 u_1(t)\Bigr)\ dt.\end{equation}

\begin{quote}{\it  If   $\kappa_2 u_2(r_0)\geq \kappa_1 u_1(r_0)$, then 
$v_2(r_0)<v_1(r_0)$.}
\end{quote}

On the contrary case, that is, if $v_1(r_0)\leq v_2(r_0)$, then $0<u_1'(r_0)<u_2'(r_0)$. 
The positivity of $\kappa_i$ and  the fact that $\kappa_1<\kappa_2$ implies
 $\kappa_1 u_1'(r_0)<\kappa_2 u_2'(r_0)$. Then $\kappa_1 u_1<\kappa_2 u_2$ in certain 
interval $(r_0,r_0+\delta)$. Let $r_1\in (r_0,R]$ be the largest number where such inequality holds. 
From (\ref{mono1}), for each $r_0<s\leq r_1$,  $v_2(s)>v_1(s)$. Thus $u_2'(s)>u_1'(s)$ and 
$\kappa_2 u_2'> \kappa_1 u_1'$. This implies $\kappa_2 u_2>\kappa_1 u_1$ for each $r_0<s\leq r_1$. 
Since $r_1$ is maximal, then $r_1=R$. 
We put now $s=R$ in (\ref{mono2})  and we obtain a contradiction. This proves the Claim.

Let us prove now the Theorem and we begin with the item 2. Assume there exists $r_0$, $0<r_0<R$, such that $u_1'(r_0)\leq u_2'(r_0)$. 
Then $v_1(r_0)\leq v_2(r_0)$. By the Claim,  $\kappa_2 u_2(r_0)<\kappa_1 u_1(r_0)$. Then 
(\ref{rotational1}) implies $(r v_2)'(r_0)<(r v_1)'(r_0)$. For certain neighbourhood on the left of $r_0$, we obtain then
$$0\leq r_0(v_2(r_0)-v_1(r_0))<r(v_2(r)-v_1(r))$$
 which it yields $v_2(r)>v_1(r)$. As 
$v_1(0)=v_2(0)=0$, there exists a last number $r_1$, $0\leq r_1<r_0$, such that 
$v_2>v_1$ in the interval $(r_1,r_0)$ and $v_2(r_1)=v_1(r_1)=0$. The Claim implies now 
$\kappa_2 u_2(r)<\kappa_1 u_1(r)$, for $r_1<r\leq r_0$. But (\ref{mono2}) yields 
$v_2(r)<v_1(r)$ and that is a contradiction. Consequently,  $u_2'<u_1'$ in $(0,R)$.

Let us prove the item 1. We use again (see the proof of  Theorem \ref{t-exi})
$$\lim_{r\rightarrow 0} \frac{v_i(r)}{r}=\kappa_i\frac{u_i(0)}{2}.$$
Because $v_1>v_2$, we infer then $u_2(0)<u_1(0)$. As $u_2'<u_1'$, an integration leads to $u_2<u_1$ in the interval $[0,R]$.

\hfill{$q.e.d.$}
\end{proof}

As conclusion, we obtain the next inclusion property:

\begin{corollary} Let $X_1$ and $X_2$ be two sessile liquid drops in 
Minkowski space $\l^3$ supported in the same spacelike plane $\Pi$. Assume that the materials of
each liquid  are different. Let $\kappa_1$ and $\kappa_2$ be the corresponding constants of capillarity. Assume that 
 the angles of contact with $\Pi$ are the same. If $\kappa_1<\kappa_2$, then it is possible to move 
$X_2$ by parallel translations  to $\Pi$ such that $X_2$ lies completely included in $X_1$.
\end{corollary}

See Remark \ref{pendent-mono}  about what it happens in 
the case of that the capillarity constant is negative.

%%%%%%%%%%%%%%%%%%%%%%
\section{Sessile liquid drops I}\label{cap-sessile1}
%%%%%%%%%%%%%%%%%%%%%%%%%

The next two sections are devoted to do a better understanding of the qualitative
properties of the shapes that adopt a sessile liquid drop, that is, when
the capillary constant $\kappa$ is positive in the Euler equation (\ref{rotational1}). In this section 
we ask as the behaviour of the profile curve, as well as, the problem of existence of sessile liquid drops. 
We remark in contrast to the Euclidean ambient, that   by Theorem \ref{t-exi} the maximal interval for  solutions of 
(\ref{Bond}) is $[0,\infty)$.

After a vertical translation, 
we consider equation
\begin{equation}\label{Bond}
\left(\frac{r u'}{\sqrt{1-u'^2}}\right)'=\kappa\ r u,\hspace*{1cm}0\leq r\leq R,
\end{equation}
 with boundary conditions 
\begin{equation}\label{ivp-3}
u'(0^+)=0\hspace*{1cm}u'(R^-)=\tanh\beta.
\end{equation}
Again, we study the initial value problem of equation (\ref{Bond}) with  boundary data 
\begin{equation}\label{Bond2}
u(0^+)=u_0,\hspace*{1cm}u'(0^+)=0.
\end{equation}
 and denote $u=u(r;u_0)$  to remark the dependence on $u_0$.  If $u_0=0$, then $u=0$, which is the 
solution associated to (\ref{Bond})-(\ref{ivp-3}) for $\beta=0$. Assume then $u_0\not=0$. 
According to the property $u(r;-u_0)=-u(r;u_0)$,   we assume $u_0>0$. Put
$$\sinh\psi=\frac{u'}{\sqrt{1-u'^2}},\hspace*{1cm}\cosh\psi=\frac{1}{\sqrt{1-u'^2}},$$
where $\psi$ is the hyperbolic tha makes $u(r)$ with the horizontal line at each point  $(r,u(r))$.
 Then equation (\ref{Bond}) writes as 
\begin{equation}\label{B-1}
(r\sinh\psi)'=\kappa \ r u
\end{equation}
or equivalently 
\begin{equation}\label{B-11}
 \sinh\psi=\frac{1}{r}\int_0^r \kappa\  t u(t)\ dt.
\end{equation}
As $u_0>0$, the integral in (\ref{B-11})  is positive, that is, $\sinh\psi>0$, and so, $u'(r)>0$ for any $r>0$.
This means that $u$ is strictly increasing in $[0,\infty)$. Then we bound the integrand in (\ref{B-11}) 
by $u_0<u(t)<u(r)$ obtaining 
\begin{equation}\label{B-2}
\frac{ \kappa   u_0}{2}<\frac{\sinh\psi(r)}{r}<\frac{\kappa  u(r)}{2}.
\end{equation}
In particular, $\sinh\psi(r)>\kappa r u_0/2\rightarrow \infty$ as $r\rightarrow\infty$. Thus, 
$$\lim_{r\rightarrow\infty}u'(r)=\lim_{r\rightarrow\infty}\frac{\sinh\psi}{\sqrt{1+\sinh^2\psi}}=1.$$
On the other hand,  (\ref{B-1}) becomes 
\begin{equation}\label{laplace1}
\frac{\sinh\psi}{r}+(\sinh\psi)'= \kappa  u.
\end{equation}

We point out that the function $\sinh\psi/r$ is strictly increasing on $r$: a direct differentiation leads to
\begin{equation}\label{creciente}
\left(\frac{\sinh\psi}{r}\right)'=\frac{(\sinh\psi)'}{r}-\frac{\sinh\psi}{r^2}=
\frac{\kappa r u-2\sinh\psi}{r^2}>0,
\end{equation}
where have used (\ref{laplace1}) and the second inequality of (\ref{B-2}). We prove then that 
$u$ is convex: by using (\ref{creciente}) and (\ref{B-2}), we have
$$(\sinh\psi)'=\kappa u-\frac{\sinh\psi}{r}>\frac{\sinh\psi}{r}>\kappa\frac{u_0}{2}>0,$$
and then, $u''>0$. We summarize up the above properties of $u$.

\begin{theorem}  Let $u$ be the profile curve for a sessile liquid drop $X$  in Minkowski space $\l^3$. Then
\begin{enumerate}
\item The function $u$ is defined  in $[0,\infty)$.
\item $u$ is strictly increasing on $r$, with $\lim_{r\rightarrow\infty}u=\infty$.
\item $u$ is convex.
\item $\lim_{r\rightarrow \infty}u'(r;u_0)=1$.
\end{enumerate}
\end{theorem}

Finally, we prove the existence of a sessile liquid drop for prescribed angle of contact.

\begin{theorem}[Existence of sessile liquid drops]\label{sessile-ex} Let $\l^3$ be the Lorentz-Minkowski space. Let $\kappa>0$ be a constant of capillarity, and $\beta,\lambda\in \r$.
\begin{enumerate} 
\item Given $R>0$, there exists a stationary sessile liquid drop $X$ supported in a disc of radius $R$ in 
some horizontal spacelike plane $\Pi$ and such that $\beta$ is the contact angle along $\partial(\overline{X}\cap \Pi)$.
The mean curvature of a point $x\in X$ is 
$\kappa x_3(x)+\lambda$.
\item Consider  $\Pi$ a  spacelike plane. Then there exists a stationary sessile liquid drop $X$ supported on $\Pi$ that makes  
a constant angle $\beta$ along $\partial(\overline{X}\cap \Pi)$. The mean curvature of a point $x\in X$ is 
$\kappa x_3(x)+\lambda$, where $x_3$ denotes the distance to $\Pi$.

\end{enumerate}
\end{theorem}

\begin{proof} We know that the surface must be a surface of revolution with respect to an 
axis orthogonal to the plane $x_3= 0$. After isometries of $\l^3$ and a change of variables, we assume that 
$\beta\geq 0$ and $\lambda=0$ and that the surface is obtained by rotating a function $u$ that is a solution of 
(\ref{Bond})-(\ref{Bond2}) with $u_0\geq 0$. Moreover, in the item 2, we assume that $\Pi$ is a horizontal plane.

\begin{enumerate}
\item If $\beta=0$, then the solution is $u=0$. Then, we assume 
$\beta\not=0$. Then  the problem reduces to find $u_0>0$ such that $u'(R;u_0)=\tanh\beta$.
 By Theorem \ref{t-exi} we have then to study the behaviour of 
$u'(r;u_0)$ with respect to $u_0$. We prove that 
$$\lim_{u_0\rightarrow 0}u'(r;u_0)=0,\ \lim_{u_0\rightarrow \infty}u'(r;u_0)=1.$$
Because $u(r;u_0)$ is increasing on $r$, 
$$v(r;u_0)=\frac{1}{r}\int_0^r t u(t)\  dt\geq \frac{1}{r}\int_0^r t u_0\  dt=\frac{u_0 r}{2}\longrightarrow \infty$$
 if $u_0\rightarrow\infty$. 
Thus
$$u'(r;u_0)= \frac{v(r;u_0)}{\sqrt{1+v(r;u_0)^2}}\rightarrow 1,
$$ as $u_0\rightarrow\infty$.
In the same way, $v(r;u_0)\rightarrow v(r;0)=0$ by the continuity of (\ref{ivp-1})-(\ref{ivp-2}) with respect to $u_0$.
 We infer then 
$u'(r;u_0)\rightarrow 0$ if $u_0\rightarrow 0$ and thus
$$\lim_{u_0\rightarrow 0^+}u'(R;u_0)=0\hspace*{1cm} \lim_{u_0\rightarrow \infty}u'(R;u_0)=1.$$
Because $u'(R;u_0)$ in continuous at $u_0$,  given $\beta\in\r$ and since $0<\tanh\beta<1$, 	there exists a
unique solution $u(r;u_0)$  such that $u'(R;u_0)=\tanh\beta$ as was to be shown.

\item Suppose $\Pi\equiv x_3=c>0$. If $\beta=0$, then the function that we seek is $u(r;c)=c$. Assume then $\beta>0$.
Consider $u_0$ varying from $u_0=0$ to $u_0=c$. Since $u$ is increasing on $r$, 
denote $r(u_0)$ the unique positive number such that $u(r(u_0);u_0)=c$ and let us study 
$u'(r(u_0);u_0)$. Since $u(r;u_0)\rightarrow 0$ as $u_0\rightarrow 0$, we know then that if $u_0\rightarrow 0$, then $r(u_0)\rightarrow \infty$. Thus 
$$\lim_{u_0\rightarrow 0}u'(r(u_0);u_0)=\lim_{r\rightarrow \infty} u'(r(u_0)=1.$$
On the other hand, if $u_0\rightarrow c$, then $r(u_0)\rightarrow 0$ and then 
$$\lim_{u_0\rightarrow c}u'(r(u_0);u_0)=\lim_{r\rightarrow 0} u'(r(u_0)=0.$$
Consequently, there will be a unique $u_0$ such that $u'(r(u_0);u_0)= \tanh\beta$ and  $u(r(u_0),u_0) = c$.
\end{enumerate}
\end{proof}

This section ends with a result of foliations of the ambient space $\l^3$ by  (entire) sessile liquid drops. 

\begin{theorem} \label{foliation}
Fix a positive number  $\kappa>0$. Then the Lorentz-Minkowski space $\l^3$ can be foliated by 
a uniparametric family of surfaces of revolution that 
satisfy the  sessile liquid drop equation for the same constant of capillarity $\kappa$. The profiles curves 
of these surfaces, $\{u=u(r;u_0);u_0\in\r\}$, are entire solutions of the initial value problem
$$\left(\frac{r u'}{\sqrt{1-u'^2}}\right)'=\kappa r u,\ r>0,\hspace*{1cm}
u(0^+)=u_0,\ \ u'(0^+)=0.$$
\end{theorem}

\begin{proof} 
Denote $u=u(r;u_0)$ the solution of  (\ref{Bond})-(\ref{Bond2}). 
We know that if $u_0=0$, then $u(r;0)=0$. Moreover, $u(r;-u_0)=-u(r;u_0)$. Thus, 
it is suffices to  show that the plane $x_2>0$ is foliated 
by the solutions of initial value problem for $u_0>0$. Given $(a,b)\in\r^2$ with $b>0$, let 
$u=u(r;u_0)$ be a  solution $u$ of (\ref{Bond}) such  that $u(a)<b$: this is possible by taking $u_0$ sufficiently 
near to $0$. Letting $u_0\rightarrow\infty$ and by the continuity with respect to the parameter $u_0$, 
there will be a $u_0$ such that $u(a;u_0)=b$. Furthermore, we know that $u_0<b$. We will prove now  that each two solutions 
$u(r;u_0)$ and $u(r;u_1)$ have no points of intersections. This is proved in the following

\begin{quote}{\it 
If $\delta>0$, then $u(r;u_0+\delta)-\delta>u(r;u_0)$ for $0<r<\infty$.}
\end{quote}

 Define the function $u_{\delta}=u(r;u_0+\delta)$, and let $\psi^{\delta}$ be the hyperbolic angle
given by  
$(r\sinh\psi^{\delta})'=\kappa r u_{\delta}$. Then 
\begin{equation}\label{integrand}
r(\sinh\psi^{\delta}-\sinh\psi)=\kappa \int_0^r t (u_{\delta}(t)-u(t))\ dt.
\end{equation}
Since the integrand is positive near to $r=0$, there exists $\epsilon>0$ such that 
$$\sinh\psi^{\delta}(r)-\sinh\psi(r)>0 \hspace*{1cm}\mbox{in }(0,\epsilon)$$
and
$$\sinh\psi^{\delta}(0)-\sinh\psi(0)=0.$$
Thus $\psi^{\delta}>\psi$ in the interval $(0,\epsilon)$. In addition, 
$$(u_{\delta}(r)-u(r))'=\tanh\psi^{\delta}(r)-\tanh\psi(r)>0.$$
Therefore the function $u_{\delta}-u$ is strictly increasing on $r$. So,  
$u_{\delta}(r)-\delta>u(r)$. Let $r_0>\epsilon$ be the first point where $u_{\delta}(r_0)-\delta=u(r_0)$. Again 
(\ref{integrand}) yields $\sinh\psi^{\delta}(r_0)-\sinh\psi(r_0)>0$.
Then $(u_{\delta}-u)'(r_0)\leq 0$.  But this implies that $\psi^{\delta}(r_0)\leq \psi(r_0)$, which  is a contradiction. 
As conclusion, $u_{\delta}-\delta>u$ in $(0,\infty)$ and this shows the Claim.

Geometrically the Claim means that  we can move $u(r;u_0+\delta)$ downwards until it touches $u$ at the point $(0,u_0)$ and 
then, the curve $u(r;u_0+\delta)$ lies completely above $u$.

\hfill{$q.e.d.$}
\end{proof}
\begin{remark} It is worth to point out that in relativity theory 
there is interest of finding 
 real-valued functions on a given spacetime, all of whose level sets provide a global time coordinate.
References can be found in review papers such as \cite{cy,mt}. In our case, this function is the 
mean curvature of entire solutions of the Euler equation for  rotational symmetric sessile liquid drops.
\end{remark}

%%%%%%%%%%%%%%%%%%%%%%%%%
\section{Sessile liquid drops II}\label{cap-sessile2}
%%%%%%%%%%%%%%%%%%%%%%%%
  
This section is devoted to obtain estimates of the size for a stationary sessile liquid drop, such that as its height and
the amount of liquid enclosed by the drop. In particular, 
 we shall show the classical estimates obtained by 
Laplace for drops in Euclidean space. See \cite{fi1} for historical comments.

Let $u$ be a solution of (\ref{Bond})-(\ref{Bond2}) with $u_0>0$. 
Moreover, we assume that $u'(R)=\tanh\beta$, that is, the liquid drop determined by $u$ 
makes an contact angle $\beta$ with the plane $\Pi\equiv \{x_3=u(R)\}$. Denote $k_m$ and $k_l$  the principal curvatures of the surface 
determined by $u$  along the meridian and latitude respectively.
Then 
$$k_m+k_l=\frac{u''}{(1-u'^2)^{3/2}}+\frac{u'}{r\sqrt{1-u'^2}}.$$
By comparing with Equation (\ref{laplace1}), 
$$k_m=(\sinh\psi)',\hspace*{1cm} k_l=\frac{\sinh\psi}{r}$$
and 
\begin{equation}\label{B-3}
k_l+k_m=\kappa u.
\end{equation}
Following \cite{fi2}, we obtain that  {\it the functions $k_m$ and $k_l$ are strictly increasing on $r$}. 
For $k_l$ is a consequence of (\ref{creciente}). For the meridian curvature function
$k_m$, we use  (\ref{B-2}) again, 
$$(k_m)'=(\sinh\psi)''=\kappa u'-\frac{\kappa u}{r}+2\frac{\sinh\psi}{r^2}>
\kappa u'-\frac{\kappa}{r}(u-u_0)=\frac{\kappa}{r}\int_0^r t u''\ dt>0.$$

Letting $r\rightarrow 0$,  $u\rightarrow u_0$ and from (\ref{B-2}),  $(\sinh\psi)/r\rightarrow
 \kappa    u_0/2$. As  consequence and from (\ref{B-3}), 
\begin{equation}\label{sectional}
k_m(0)= \kappa   u_0- \frac{\kappa   u_0}{2}= \frac{\kappa    u_0}{2},\hspace*{1cm}k_l(0)= \frac{\kappa    u_0}{2}.
\end{equation}
We compare our liquid drop with pieces of hyperbolic planes. After a isometry of $\l^3$, 
 hyperbolic planes are defined by 
$$y(r)=\sqrt{r^2+\mu^2}+c, \hspace*{1cm}\mu>0, c\in\r,$$
with mean curvature $H=1/\mu$ with respect to the future-directed orientation. 

We consider  $\Sigma^1$ and $\Sigma^2$  two hyperbolic caps obtained by rotation of functions  $y_1$ and $y_2$
defined in the interval $[0,R]$  as follows. The surface $\Sigma^1$ has  mean curvature $H_1=\kappa   u_0/2$
and  $y_1(0)=u_0$. The curve $y_2$ satisfies $y_2(0)=u_0$, and  $y_2'(R)=u'(R)$. 
Then 
$$ \mu_1=\frac{2}{\kappa  u_0},\hspace*{1cm} \mu_2=R\frac{\sqrt{1-u'(R)^2}}{u'(R)}=\frac{R}{\sinh\beta}, 
\hspace*{1cm}c_i=u_0-\mu_i. $$
We shall prove that 

\begin{quote} {\it The sessile liquid  drop defined by $u$ is bounded by 
the hyperbolic caps $\Sigma^1$ and  $\Sigma^2$.}
\end{quote}

\begin{figure}[htbp]
\includegraphics[width=11cm,height=6.4cm]{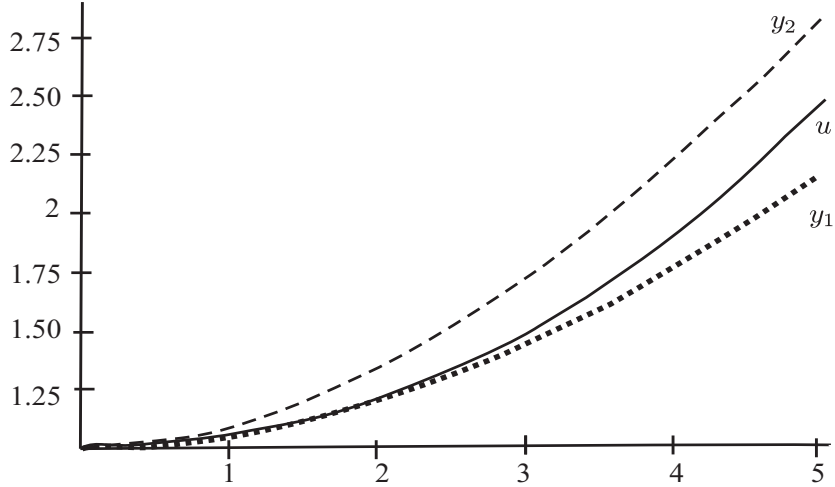}
\caption{Comparison between the profiles $u$, $y_1$ and $y_2$. Here $\kappa=0.2$ and $u_0=1$.}
\label{fig1}
\end{figure}

See Figure \ref{fig1}. The  constant sectional curvature $k_1$ of $\Sigma^1$ is $H_1$ and 
$k_1(r)=k_1(0)=k_m(0)<k_m(r)$. Since $k_m=(\sinh\psi)'$, 
$$\left(\frac{y_1'}{\sqrt{1-y_1'^2}}\right)'<\left(\frac{u'}{\sqrt{1-u'^2}}\right)'.$$
Integrating this expression and since $y_1$ and $u$ have the same initial conditions, we obtain
$y_1(r)<u(r)$ for any $r>0$.

On the other hand, $y_2'(R)=u'(R)>y_1'(R)$, and  the sectional curvature $k_2$  of 
$\Sigma^2$ satisfies  $k_2(r)>k_1(r)$ for any $r$. 
From (\ref{sectional}), $k_m<k_2$ in some interval $(0,\epsilon)$ on the right of $r=0$. 
Because the initial conditions are the same  for $u$ and $y_2$, we know that $u<y_2$ in some interval $(0,\epsilon)$. 
We assert that $u<y_2$ on $(0,R]$. The next reasoning is similar as in
Theorem \ref{sessile-mono}. On  the contrary case, let $\bar{r}<R$ be the first point such that 
$u(\bar{r})=y_2(\bar{r})$. Then  necessarily, 
$k_2(\bar{r})\leq k_m(\bar{r})$ and $y_2'(\bar{r})\leq u'(\bar{r})$ and so, 
$\psi_2(\bar{r})\leq \psi(\bar{r})$ (if not, it holds still $u<y_2$ at $r=\bar{r}$). 
As $u'(0)=y_2'(0)$, we have $\psi(0)=\psi_2(0)$ and 
$$\int_0^{\bar{r}}(k_m-k_2)\ dr=\sinh(\psi){\big|}_0^{\bar{r}}-\sinh({\psi}_{2}){\big|}_0^{\bar{r}}\geq 0.$$
As $k_m$ is strictly increasing, $k_m>k_2$ in $(\bar{r},R]$ and therefore
$$\int_{\bar{r}}^R(k_m-k_2)\ dr >0.$$
Consequently $\int_0^R (k_m-k_2)\ dr>0$, but 
$$\int_0^R (k_m-k_2)\ dr =\sinh(\psi(R))-\sinh(\psi_2(R))=0$$
since $u'(R)=y_2'(R)$, which finishes the proof of the Claim.

This leads to a bound of the 
volume of the capillary drop.  According to (\ref{B-11}), the volume $V$ of the liquid drop $u$ that 
is solution of (\ref{Bond})-(\ref{Bond2}) is
\begin{equation}\label{volume}
V=2\pi\int_0^R r u(r)\ dr=\frac{2\pi R}{\kappa }\sinh\beta,
\end{equation}
where $u'(R)=\tanh\beta$. By $V$ we mean the volume of the cone defined by the graph of $u$ but not  the physical volume 
${\cal V}$ enclosed by the generated drop by $u$ and the plane $x_3=u(R)$ (later, we shall obtain bounds for the 
volume ${\cal V}$).
For the computation of the volume $V_i$ of each surface $\Sigma^i$
between $r=0$ and $r=R$, we define 
$$F(u_0;s)= \frac{R^2}{2} (u_0-s)+\frac13\left((s^2+R^2)^{3/2}-s^3\right).$$
Then it is straightforward that   $V_i=2 \pi F(u_0;\mu_i)$, $i=1,2$. As the surface defined by $u$ lies between $\Sigma^1$ and $\Sigma^2$, $V_1<V<V_2$. 
This enables us to estimate the value of $u(0)=u_0$ and $u_R:=u(R)$  in terms of the contact angle $\beta$ at $r=R$.

\begin{theorem} Let $u$ be a solution of (\ref{Bond})-(\ref{Bond2}). Then $u_0$ and $u_R$ 
satisfy
\begin{equation}\label{esti-u0}
\frac{2\sinh\beta}{R\kappa }+\frac{R}{\sinh\beta}+\frac{2R}{3}\frac{1-\cosh^3\beta}{\sinh^3\beta}<u_0<\frac{2\sinh\beta}{R \kappa }.
\end{equation}
\begin{equation}\label{esti-u1}
u_R<\frac{ 2\sinh\beta}{R \kappa}+R\frac{\cosh\beta}{\sinh\beta}+\frac{2 R}{3}\frac{1-\cosh^3\beta}{\sinh^3\beta}.
\end{equation}
\end{theorem}

\begin{proof}
From inequality 
$V<V_2$ we obtain the lower bound for $u_0$ in (\ref{esti-u0}). For the other estimate, we 
know that $V_1<V$, and so $F(u_0;\mu_1)<(R\sinh\beta)/\kappa$. Since $\partial F/\partial u_0=R^2/2>0$,
let $u_0^+$ be the unique solution of 
$(R\sinh\beta)/\kappa =F(u_0^+;\mu_1)$. By the monotonicity of $F$ with 
respect to $u_0$,  we have $u_0<u_0^+$.  Moreover, 
$$F\left(\frac{2\sinh\beta}{R\kappa} ;\mu_1\right)>\frac{R\sinh\beta}{\kappa} =F(u_0^+;\mu_1)$$
 and the monotonicity of $F$ implies again 
$u_0^+<(2\sinh\beta)/(R \kappa)$, proving the upper bound in (\ref{esti-u0}).

For the estimate of $u_R$, we descend now the hyperbolic cap $\Sigma^2$ until $y_2$ touches the drop $u$ at the point 
$(R,u_R)$, where both function $y_2$ and $u$ are tangent. The principal curvature of $u$ for the meridian
at $r=R$ is $k_m(R)=\kappa u_R-\sinh\beta/R$. By using (\ref{B-2}), 
$$k_m(R)>\frac{\sinh\beta}{R}=H_2=k_2.$$
A similar reasoning as in the proof of the above Claim shows that   $\Sigma^2$ lies completely below $M$. 
The estimate (\ref{esti-u1})  follows now  by comparing the volumes of $\Sigma^2$ and $M$.

\hfill{$q.e.d$} 
\end{proof}
Inequality (\ref{esti-u0}) is well known in Euclidean space and was obtained by Laplace \cite{la}.
See also (\cite{fi,fi2,fi3}).  The above estimates (\ref{esti-u0})-(\ref{esti-u1})
for $u_0$ and $u_R$ give an immediate bound for  the height of a liquid drop, independently of
the constant of capillarity $\kappa$.

\begin{corollary} \label{sessile-co} 
Consider $X$ a stationary sessile liquid drop in $\l^3$ making a hyperbolic angle $\beta$ with the support plane $\Pi$.
Then the height $q$ of $X$ with respect to $\Pi$  satisfies
$$q<R\frac{\cosh\beta-1}{\sinh\beta},$$
where $R$ is the radius of the disc that the drop wets on $\Pi$.
\end{corollary}

Again, we consider equation (\ref{Bond}). Since the solution $u$ 
is strictly increasing on $r$ and $u'(r)=\tanh\psi$, we can take $\psi$ as a new variable. This leads the following system equivalent to Equation (\ref{Bond}) of two first differential 
equations for $r$ and $u$ defined by:
\begin{equation}\label{system}
r'(\psi)=\frac{r\cos\psi}{\kappa r u-\sinh\psi},\hspace*{1cm}u'(\psi)=\frac{r\sinh\psi}{\kappa r u-\sinh\psi}
\end{equation}
For the  estimates of the volume that we seek, we need the following

\begin{theorem} \label{sessile-II}
Let $u$ be  a solution  of  the initial value problem (\ref{Bond})-(\ref{Bond2}). Then 
 \begin{equation}\label{cota-1}
\frac{u_0}{2}+\sqrt{\frac{u_0^2}{4}+\frac{2}{\kappa}(\cosh\psi-1)}<
u(\psi)<\sqrt{\frac{4}{\kappa}(\cosh\psi-1)+u_0^2}
\end{equation}
As a consequence, 
 \begin{equation}\frac{2}{\kappa}(\cosh\psi-1)<u^2-u_0^2<\frac{4}{\kappa}(\cosh\psi-1).
\end{equation}
\end{theorem}

\begin{proof}
From (\ref{B-2}) we know 
$$\frac{ \kappa r u}{2}< \kappa r u-\sinh\psi< \kappa r u-\frac{ \kappa r u_0}{2},$$
and substituting into the value of $u'(\psi)$ in (\ref{system}), we obtain
$$\frac{r \sinh\psi}{ \kappa r u- \kappa r u_0/2}<u'(\psi)<\frac{r\sinh\psi}{\kappa r u /2}.$$
Integrating both inequalities between $0$ and $\psi$, we obtain (\ref{cota-1}). 

\hfill{$q.e.d$}
\end{proof}
We point out that in these inequalities do not appear the variable $r$. In contrast to it happens in the Euclidean ambient, (\ref{cota-1}) hold for any value of $\psi$.  Theorem \ref{sessile-II} allows to estimate the height of a sessile liquid drop. 
According to (\ref{esti-u1}), let us denote
$${\cal F}(\kappa,R,\beta)=\frac{ 2\sinh\beta}{R \kappa}+R\frac{\cosh\beta}{\sinh\beta}+\frac{2 R}{3}\frac{1-\cosh^3\beta}{\sinh^3\beta}.$$
We improve Corollary \ref{sessile-co}. 

\begin{corollary}  Let $u$ be  a solution of (\ref{Bond})-(\ref{Bond2}). 
Assume that $u'(R)=\tanh\beta$. Denote by $q$ the height of a solution $u$,
$q=u_R-u_0$. Then 
\begin{equation}\label{esti-q}
\frac{2(\cosh\beta-1)}{\kappa {\cal F}(\kappa,R,\beta)}<
q<R\frac{\cosh\beta-1}{\sinh\beta}.
\end{equation}
\end{corollary}

\begin{proof} Using twice the inequality (\ref{cota-1}), we obtain
\begin{eqnarray*}
u_R-u_0&>&-\frac{u_0}{2}+\sqrt{\frac{\kappa}{2}(\cosh\beta-1)+\frac{u_0^2}{4}}\\
&=& \frac{2}{\kappa}(\cosh\beta-1)\frac{1}{\frac{u_0}{2}+\sqrt{\frac{\kappa}{2}(\cosh\beta-1)+\frac{u_0^2}{4}}}\\
&>&\frac{2(\cosh\beta-1)}{\kappa u_R}.
\end{eqnarray*}
We conclude then the lower bound in (\ref{esti-q}) using  that $u_R<{\cal F}(\kappa,R,\beta)$.

\hfill{$q.e.d$}
\end{proof}
We are in condition to obtain lower and upper bounds of the volume ${\cal V}$ determined by  a sessile liquid drop and 
 the plate $\Pi$.

\begin{theorem}\label{estima-vo} Let $X$ be a sessile liquid drop in Minkowski spec $\l^3$ and supported in a 
spacelike plane $\Pi$. Denote $R$ and $\beta$  respectively 
the radius of the disc that $\overline{X}$ wets in $\Pi$ and  the hyperbolic angle of 
contact that makes $\overline{X}$ with $\Pi$ along the liquid-air-solid interface $\partial(\overline{X}\cap\Pi)$. Then the volume 
of liquid ${\cal V}$ is estimated by 
\begin{equation}\label{esti-v}\frac{\pi}{3} C(\beta)\left(\frac{2}{\kappa {\cal F}(\kappa,R,\beta)}\right)^3\leq{\cal V}
\leq \frac{\pi R^3}{3} \frac{C(\beta)}{\sinh^3\beta},
\end{equation}
where $C(\beta)=\cosh^3\beta-3\cosh\beta+2$.
\end{theorem}

\begin{proof} Without loss of generality, we assume that $X$ is determined by  a solution  $u$ of  equation (\ref{Bond})-(\ref{Bond2}). Assume that $u$ intersects a horizontal spacelike plane $\Pi$ at $(R,u_R)$ 
with $u'(R)=\tanh\beta$. The volume that encloses $X$ with $\Pi$ is 
$${\cal V}=\pi R^2 u_R-\frac{2\pi R\sinh\beta}{\kappa}.$$
The upper estimate is a consequence of the upper bound of $u_R$ in (\ref{esti-u1}). 
For the other one, we consider the hyperbolic cap $\Sigma^3$ obtained by rotating a function $y_3$ such that 
$y_3$ is  tangent to $u$ at $(0,u_0)$ and the  mean curvature of $\Sigma^3$ is $H=\kappa u_R/2$.
When $y_3$ writes in polar coordinates, that is, 
$r=m\sinh\phi$ and $v=u_0+m(\cosh\phi-1)$, $m=1/H$, then $\phi$ satisfies 
$$(r\sinh\phi)'=\kappa  r u_R.$$
Because $u$ is strictly increasing, in the interval $(0,R]$ we have
$$r(\sinh\phi-\sinh\psi)=\kappa\int_0^r t\ (u_R-u)\ dt>0.$$
Consequently $\phi(r)>\psi(r)$ and $y_3'(r)=\tanh\phi(r)>\tanh\psi(r)=u'(r)$. Because 
$y_3$ and $u$ have the same initial condition at $r=0$, then $y_3<u$ in $(0,R]$.
We prove now that at the point
 $r^\beta$ where $y_3'(r^{\beta})=u'(R)$, it holds $y_3(r^{\beta})<u_R$.
Now  (\ref{esti-q}) yields 
$$u_R-u_0>\frac{2(\cosh\beta-1)}{\kappa u_R}=y_3(r^{\beta})-u_0.$$
  It follows that  the enclosed volume by $y_3$ between in $[0,r^\beta]$ is a lower bound of ${\cal V}$. A computation of this 
volume gives 
$${\cal V}\geq \frac{\pi}{3} C(\beta)\left(\frac{2}{\kappa u_R}\right)^3$$
and finally we use $u_R<{\cal F}(\kappa,R,\beta)$ again. 

\hfill{$q.e.d$}
\end{proof}

We follows obtaining estimates by using Theorem \ref{sessile-II}. The next two results  give an upper and lower  bounds of 
the value $u$ in terms of the hyperbolic angle $\psi$ of contact with the plane $x_3=u$. 

\begin{theorem}\label{sup} Let $u$ be a solution of (\ref{Bond})-(\ref{Bond2}). Then
\begin{equation}\label{cota-sup}
u(\psi)<\frac{\sinh\psi}{r \kappa}+\sqrt{\frac{2}{\kappa}(\cosh\psi-1)+\left(\frac{\sinh\psi}{\kappa r}\right)^2}.
\end{equation}
\end{theorem}

\begin{proof}
The expression of $u'(\psi)$ in (\ref{system}) leads also
$$(\kappa u-\frac{\sinh\psi}{r}) u'=\sinh\psi.$$
We do two integrations: first, between $u=0$ and a value $u=\lambda$; and second, between $u=\lambda$
and $u=u(\psi)$:
\begin{equation}\label{doble}
\cosh\psi-1=\int_{u_0}^{\lambda} (\kappa u-\frac{\sinh\psi}{r}) du+\int_{\lambda}^{u}
(\kappa u-\frac{\sinh\psi}{r})\ du.
\end{equation}
Let us use (\ref{B-2}).
For the first integral, we bound from below by $\kappa u u'/2$; in the second one, the
summand $\sinh\psi/r$ is bounded by its value at $r$ (remembering that $\sinh\psi/r$ is increasing on $r$, see 
(\ref{creciente})). Then
$$\cosh\psi-1>\frac{\kappa}{4}(\lambda^2-u_0^2)+
\frac{\kappa}{2}(u^2-\lambda^2)-\frac{\sinh\psi}{r} (u-\lambda),$$
Then we obtain
$$u(r)<\frac{\sinh\psi}{r \kappa}+\sqrt{\frac{2}{\kappa}(\cosh\psi-1)+\left(\lambda-\frac{\sinh\psi}{\kappa r}\right)^2-
\left(\frac{\lambda^2-u_0^2}{2}\right)}.$$
The radicand  
attains a minimum at $\lambda_0=2\sinh\psi/(\kappa r)$. Recall that from 
(\ref{B-2}), the value $\lambda_0$ of $u$ lies in the interval of integration. Thus
$$u(r)<\frac{\sinh\psi}{r \kappa}+\sqrt{\varphi(\lambda_0)}=
\frac{\sinh\psi}{r \kappa}+\sqrt{\frac{2}{\kappa}(\cosh\psi-1)+\frac{u_0^2}{2}-\left(\frac{\sinh\psi}{\kappa r}\right)^2}.$$
Using (\ref{B-2}),  $\sinh\psi/r>\kappa u_0/2$ and this concludes the proof. 

\hfill{$q.e.d$}
\end{proof} 
Denote 
$$\theta=\frac{r(\psi)}{\cosh(\psi/2)},\hspace*{1cm}p=\sqrt{1+\kappa\theta^2}.$$

\begin{theorem} \label{inf}Let $u$ be a solution of (\ref{Bond})-(\ref{Bond2}). Then
\begin{equation}\label{cota-inf}
u(\psi)>\sqrt{\frac{1+p}{p}}\sqrt{\frac{2(\cosh\psi-1)}{\kappa}+\frac{u_0^2}{4}(1+p)}.
\end{equation}
\end{theorem}

\begin{proof} 
Inequality (\ref{cota-sup})  writes as 
\begin{equation}\label{2.53}
u(\psi)<\frac{\sinh\psi}{\kappa r}(1+p).
\end{equation}
Take  (\ref{doble}) again, but we bound it from below. In the 
first integral, we consider $\sinh\psi/r>\kappa u_0/2$. In the second one, we use  (\ref{2.53}) 
to obtain 
$$\kappa u-\frac{\sinh\psi}{r}<\kappa u\frac{p}{1+p}.$$
Then (\ref{doble}) yields
$$\frac{2(\cosh\psi-1)}{\kappa}<\lambda^2-\lambda u_0 +(u(\psi)^2-\lambda^2)\frac{p}{1+p}.$$
This function attains its minimum at $\lambda_1=u_0 (1+p)/2$. 

\begin{quote} {\it  The value $\lambda_1$ lies in the range of the function $u$.}
\end{quote}

 As $p>1$, we know then 
$u_0<u_0 (1+p)/2$. From (\ref{B-2}) we have $\kappa^2u_0^2<4\sinh^2\psi/r^2$. Then
\begin{eqnarray*}\left(\frac{u_0 p}{2}\right)^2&=&\frac{u_0^2}{4}+\frac{\kappa u_0^2}{4}\left(\frac{r}{\cosh(\psi/2)}\right)^2<
\frac{u_0^2}{4}+\frac{\sinh^2\psi}{\kappa\cosh^2(\psi/2)}\\
&=&\frac{u_0^2}{4}+\frac{2(\cosh\psi-1)}{\kappa} .
\end{eqnarray*}
The proof of the Claim ends by using  (\ref{cota-1}).

As conclusion, in the minimum $u=\lambda_1$, we obtain
$$\frac{2(\cosh\psi-1)}{\kappa}<u^2\frac{p}{1+p}-\frac{u_0^2}{4}(1+p),$$
showing (\ref{cota-inf}). 

\hfill{$q.e.d$}
\end{proof}
The last estimate that we shall obtain gives a lower bound for $u$ at $r=0$.

\begin{theorem} Let $u$ be a solution of (\ref{Bond})-(\ref{Bond2}). For any value of $r$, there holds
\begin{equation}\label{u0}
u_0>\frac{\sinh\psi(r)}{\kappa r}\frac{1+\sqrt{1+\kappa r^2}}{e^{-1+\sqrt{1+\kappa r^2}}}
\end{equation}
\end{theorem}

\begin{proof} For the value of $r'(\psi)$ in (\ref{system}), we consider the 
inequality (\ref{2.53}) obtaining
$$\frac{p}{r} r'(\psi)>\coth\psi .$$
We  bound $\cosh(\psi/2)$ by $1$ in the function $p$. Then 
$$\frac{\sqrt{1+\kappa r^2}}{r}r'(\psi)>\coth\psi.$$ 
Let us integrate this inequality between $\psi=0$ and $\psi$. By (\ref{B-2}), we know that 
$$\lim_{\psi\rightarrow 0} \frac{\sinh\psi}{r}=\frac{\kappa u_0}{2}.$$
This leads to the result stated in (\ref{u0}). 

\hfill{$q.e.d$}
\end{proof}
We prove now that for each volume there exists a unique sessile liquid drop.

\begin{theorem}\label{sessile-vo} Let $\kappa>0$ be  a constant of capillarity. 
Then for each  $V>0$ and $\beta\in\r$, there exists a unique (up isometries)
sessile liquid drop $X$ in Minkowski space $\l^3$ supported in a spacelike plane  $\Pi$ such that 
the volume of the drop is $V$  and $\beta$ is the boundary angle of contact between $X$ 
and $\Pi$ along $\partial (\overline{X}\cap\Pi)$.
\end{theorem}

 \begin{proof} We follow the same ideas than in Euclidean space \cite{fi}. 
We know that the liquid 
drop is described by a surface of revolution whose profile curve $u$ is a solution 
 of Equation (\ref{rotational1}). Since vertical translations do not change the  enclosed volume and the contact angle, 
 we can suppose that $\lambda=0$. We may assume without loss of generality that  
$\beta$ and $u_0$ are positive  and the drop touches $\Pi$ at $r=R$. Then $\Pi\equiv\{x_3=u_R\}$ and $u'(R)=\tanh\beta$. By using 
(\ref{volume}), the volume ${\cal V}$ of the sessile liquid drop
determined by $u$ is 
$${\cal V}=\pi R^2 u_R-\frac{2\pi R\sinh\beta}{\kappa}.$$
We study the behaviour of ${\cal V}$ according to the variable $u_0$. Note that as in 
Euclidean ambient,  ${\cal V}=
{\cal V}(u_0)$ is continously differentiable function on $u_0$.  

For the existence, we take  limits of ${\cal V}$ when $u_0$ goes to $\infty$ and $0$. 
By (\ref{B-2}), 
 $\kappa u_0/2<\sinh\beta/R$. Thus $R=R(u_0)\rightarrow 0$ if $u_0\rightarrow \infty$.
Because $u$ is increasing on $r$,  ${\cal V}<\pi R^2(u_R-u_0)$. On the other hand, and since $u$ is convex, the number ${\cal V}$ is less than  the volume of the 
surface of revolution generated by the segment of the  straight-line that joins the points $(0,u_0)$ and 
$(R,u_R)$. Then 
$${\cal V}>\frac{\pi}{3}R^2 (u_R-u_0).$$
It follow  from (\ref{esti-q}) that 
$$\pi R^2 (u_R-u_0)<\pi R^3\frac{\cosh\beta-1}{\sinh\beta}\longrightarrow 0,$$
as $u_0\rightarrow \infty$. We consider now the behaviour of ${\cal V}$ as $u_0\rightarrow 0$. We point out that 
the estimate (\ref{u0}) yields $R\rightarrow \infty$. By using  (\ref{esti-q}) and (\ref{cota-sup}), we obtain 
$$\frac{\pi}{3}R^2 (u_R-u_0)>\frac{\pi}{3}R^2\ \frac{2(\cosh\beta-1)}{\kappa \alpha(R)},$$
where 
$$\alpha(R)=\frac{\sinh\beta}{\kappa R}+\sqrt{\frac{2}{\kappa}(\cosh\beta-1)+\left(\frac{\sinh\beta}{\kappa R}\right)^2}.$$
But
$$\lim_{R\rightarrow\infty}\frac{R^2}{\alpha(R)}=\infty.$$
Thus, for fixed $\beta$, every value $V$ between $0$ and $\infty $ is attained by ${\cal V}$ and this proves the existence.

The uniqueness of $u$ is showed in proving that 
$$\stackrel{\cdot}{\cal V}=\frac{\partial {\cal V}}{\partial u_0}<0.$$
The proof is similar as in \cite{fi} and the details are omitted. 

\hfill{$q.e.d.$}
\end{proof}

Some of the estimates obtained in this section are compared with numerical calculations in Table \ref{table1}. 
We compute
the height $q$ and volume ${\cal V}$ for different values of $\kappa$ and 
let us compare with the  estimates in  (\ref{esti-q}) and (\ref{esti-v}) respectively. 
On the other hand, Table \ref{table2}contains computed data comparing the value of $u$ with the lower and upper bounds obtained in 
Theorems \ref{cota-sup} and \ref{cota-inf}.

\begin{table}
\begin{tabular}{||c||l|l|l|l|l|l||}\hline\label{table1}
$\kappa$&$u_0$ & $\beta$& $q$ & $R \frac{\cosh\beta-1}{\sinh\beta}$ &  ${\cal V}$ & $\frac{\pi R^3}{3}\frac{C(\beta)}{\sinh^3\beta}$ \\ \hline\hline
& 1&  1.81411   &1.82968 & 2.15915 & 23.7166 & 25.2538\\ \cline{2-7}
& 2&  2.31026    &2.25872&2.45834&26.3753&26.9749\\ \cline{2-7}
$1$&3&2.60668    &2.45583&2.58774&27.2134&27.5101\\ \cline{2-7}
& 4&   2.82388 &2.57013&2.66372&27.592&27.7613\\ \cline{2-7}
&5& 2.99749   &2.64488&2.71476&27.7971&27.9031\\ \cline{2-7}\hline\hline
& 1& 2.65792   &2.31749&2.60698&26.9017&27.5782\\ \cline{2-7}
& 2&  3.07744    &2.59049&2.73571&27.7404&27.9549\\ \cline{2-7}
$2$&3&  3.34433&2.70722&2.79551&27.9843&28.0818\\ \cline{2-7}
& 4& 3.54684   &2.77236&2.83195&28.0908&28.1437\\ \cline{2-7}
&5& 3.71202    &2.81392&2.85693&28.1474&28.1794\\ \cline{2-7}\hline\hline
& 1&3.12955    &2.51278&2.74857&27.631&27.9848\\ \cline{2-7}
& 2& 3.51287   &2.71508&2.82631&28.0284& 28.1349\\ \cline{2-7}
$3$&3&3.76713    &2.7988&2.86442&28.1418& 28.189\\ \cline{2-7}
& 4&3.96351    &2.84468&2.88815&28.1909&28.2161\\ \cline{2-7}
&5&  4.12532  &2.87363&2.90459&28.2168&28.2319\\ \cline{2-7}\hline\hline
& 1& 3.45367   &2.61892&2.81604&27.9035&28.1181\\ \cline{2-7}
& 2& 3.81658   &2.78087&2.87083&28.1336&28.1968\\ \cline{2-7}
$4$&3& 4.06387   &2.84645&2.89865&28.1988&28.2265\\ \cline{2-7}
& 4&   4.25719 &2.88199&2.91621&28.2268&28.2416\\ \cline{2-7}
&5& 4.41732   &2.90424&2.92846&28.2416&28.2504\\ \cline{2-7}
\hline\hline
\end{tabular}
\caption{Estimates of the height and volume for a sessile liquid drop. The value of the radius is  $R=3$}
\end{table}

\begin{table}
\begin{tabular}{||c||l|l|l|l|l||}\hline\label{table2}
$\kappa$&$u_0$ & $\beta$& Estimate \ref{cota-inf} & $u(r)$ &  Estimate \ref{cota-sup} \\ \hline\hline
& 1&  2.34282  &3.62841 & 3.79801 & 4.47819 \\ \cline{2-6}
& 2&  2.76649    &5.0108&5.2461&6.20917\\ \cline{2-6}
$1$&3&3.02607    &6.18545&6.4486&7.59829\\ \cline{2-6}
& 4&   3.22042 & 7.29241& 7.56535& 8.8557\\ \cline{2-6}
&5& 3.37825& 8.36791& 8.64145& 10.0445\\ \cline{2-6}\hline\hline
& 1& 3.18307& 4.10559& 4.31156& 5.15817\\ \cline{2-6}
& 2&  3.53125& 5.3156& 5.58775&    6.67784\\ \cline{2-6}
$2$&3&  3.76251& 6.40932& 6.70556& 7.95966\\ \cline{2-6}
& 4& 3.94344& 7.47138&  7.77123& 9.15634\\ \cline{2-6}
&5& 4.09222& 8.5138& 8.81309& 10.2938\\ \cline{2-6}\hline\hline
& 1&3.6462& 4.28634& 4.51045& 5.42549\\ \cline{2-6}
& 2& 3.96438& 5.42732& 5.71393&     6.85477\\ \cline{2-6}
$3$&3&4.18372& 6.4876& 6.79809& 8.08826\\ \cline{2-6}
& 4&4.35882& 7.53094& 7.84419& 9.25601\\ \cline{2-6}
&5& 4.50492& 8.5633& 8.87327& 10.3789\\ \cline{2-6}\hline\hline
& 1& 3.96457& 4.38184& 4.61769& 5.56912\\ \cline{2-6}
& 2& 4.26557 &5.4825& 5.78024 &  6.94216\\ \cline{2-6}
$4$ &3& 4.48006& 6.52895 &6.84606 &8.15756\\ \cline{2-6}
& 4&   4.65206 &7.56245 &  7.88172 &9.31013\\ \cline{2-6}
&5& 4.79706& 8.59055& 8.90404& 10.4283\\ \cline{2-6}
\hline\hline
\end{tabular}
\caption{Estimates of the value $u$ for sessile liquid drop according to Theorems \ref{sup} and \ref{inf}.
 Here $r=4$.}
\end{table}

%%%%%%%%%%%%%%%%%%%%%%%%
\section{Pendent liquid  drops}\label{cap-pendent}
%%%%%%%%%%%%%%%%%%%%%%%%

This last section of the present  paper is devoted to the case $\kappa<0$  in the 
Euler equation (\ref{rotational1}). We assume $\lambda=0$ and we consider  solutions of
\begin{equation}\label{rotational3}
\frac{d}{dr}\left(\frac{r u'(r)}{\sqrt{1-u'(r)^2}}\right)=\kappa\ r u(r),\ 0\leq r, \hspace*{1cm}\kappa<0.
\end{equation}
\begin{equation}\label{rotational33}
u(0^+)=u_0,\hspace*{1cm}u'(0^+)=0.
\end{equation}

\begin{figure}[htbp]
\includegraphics[width=7cm,height=2.8cm]{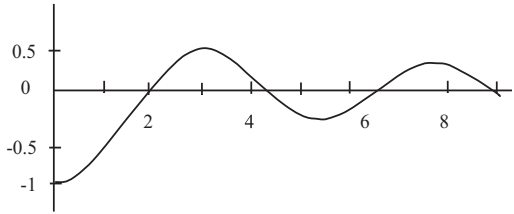}
\caption{The profile of a pendent liquid drop. Here $\kappa=-2$, $u_0=-1$.}\label{fig3}
\end{figure}

The main fact that we shall show is that a solution $u$  oscillates about the $r$-axis with successively decreasing extrema. 
See Figure \ref{fig3}.
In contrast to the Euclidean case \cite{cf,we2}, the spacelike character of our surfaces 
prohibits the presence of vertical points  for the function $u$: recall that $u'=\tanh\psi\in(-1,1)$.

From the theorem of existence and uniqueness, we know
 that (\ref{rotational3})-(\ref{rotational33}) has a unique solution for any initial data. In addition, the function
$u$ is defined for any $r$ (we emphasize  that this is a difference with the Euclidean case). If we denote 
by $u=u(r;u_0)$ the dependence on the initial value $u_0$, we have 
$u(r;-u_0)=-u(r;u_0)$. Without loss of generality, we assume $u_0\leq 0$.For $u_0=0$, the solution is $u=0$. Let now 
$u_0<0$.

Set $\sinh\psi=u'/\sqrt{1-u'^2}$.  Then equation (\ref{rotational3}) writes as 
\begin{equation}\label{pendent11}
(r\sinh\psi)'=\kappa r u.
\end{equation}
 Since
$\psi'(r)=\psi_u\tanh u$, we can write this equation as
\begin{equation}\label{pendent1}
\frac{\sinh\psi}{r}+(\sinh\psi)'=\frac{\sinh\psi}{r}+(\cosh\psi)_u=\kappa u.
\end{equation}
As in the case of sessile liquid drops, we have a similar result as in (\ref{B-2}), 

\begin{lemma} \label{PP-2} Let $u$ be a solution of (\ref{rotational3})-(\ref{rotational33}). Then there exists 
an interval $[0,R]$ where $u$ is strictly increasing and
\begin{equation}\label{P-2}
\frac{\kappa u}{2}<\frac{\sinh\psi}{r}<\frac{\kappa u_0}{2},
\end{equation}
for each $r\in (0,R)$. 

\end{lemma}
\begin{proof}
Integrating (\ref{pendent11}) between $0$ and $r$, we obtain 
$$\sinh\psi=\frac{\kappa}{r}\int_0^r t\ u(t)\ dt.$$
As $u_0 <0$, $r\sinh\psi>0$ in a neighbourhood $[0,R)$ of $r=0$. Then 
$\sinh\psi>0$ and $u'>0$ in some interval $(0,R)$ around $r=0$. This means that $u$ is strictly increasing in $[0,R)$ and $\psi>0$ in $(0,R)$. For each $r\leq R$, $u_0<u(t)<u(r)$, $t\in (0,r)$. Then in the above integral, we bound 
$t u(t)$ by $t u_0<t u(t)<t u(r)$ and this shows (\ref{P-2}).

\hfill{$q.e.d$}
\end{proof}
As conclusion , we know that $u$ is negative in some interval on the right of $r=0$. Let $r_o<R$ be a number, where 
$u([0,r_o))<0$. The next result says that such numbers $r_o$ are bounded from above.

\begin{lemma}
Let $u$ be a solution of (\ref{rotational3})-(\ref{rotational33}) and $r_o$ a real number such that $0<r_o<R$ and 
$u([0,r_o))<0$. Then for each $0<a<r_o$ there exists a constant $C(a)$ such that  $r_o<C(a)$.   
\end{lemma}

\begin{proof} From (\ref{pendent11}) and since $u<0$ in $[0,r_o)$, the function $r\sinh\psi(r)$ is increasing on $r$ 
in $[0,r_o)$. Then for each $r\in(0,r_o)$, $r\sinh\psi(r)>a\sinh\psi(a)$. In particular, $\sinh\psi>a\sinh\psi(a)/r$ and 
we put it into the expression of $u'$:
$$u'(r)=\tanh\psi=\frac{\sinh\psi}{\sqrt{1+\sinh^2\psi}}.$$
As the function $x/\sqrt{1+x^2}$ is strictly increasing, we obtain 
$$u'(r)>\frac{a\sinh\psi(a)}{\sqrt{r^2+a^2\sinh^2\psi(a)}}.$$
Integrating this inequality in the interval $[a,r_o]$, 
$$-u(a)>u(r_o)-u(a)>\int_a^{r_o}\frac{a\sinh\psi(a)}{\sqrt{r^2+a^2\sinh^2\psi(a)}}\ dr.$$
\begin{eqnarray*}
r_o&<&a\left(\cos\psi(a)\sinh\left(\frac{-u(a)}{a\sinh\psi(a)}\right)+\cosh\left(\frac{-u(a)}{a\sinh\psi(a)}\right)\right)\\
&<& C(a):=a\left(\cos\psi(a)\sinh\left(\frac{-2}{\kappa a^2}\right)+\cosh\left(\frac{-2}{\kappa a^2}\right)\right),
\end{eqnarray*}
where we have used the left inequality of (\ref{P-2}). 

\hfill{$q.e.d$}
\end{proof}
The constant $C(a)$ is minimized at $r=2/\sqrt{-\kappa}$, where $C(a)<2\sqrt{e}/\sqrt{-\kappa}$. 
We consider $R$ the (maximal) number given in Lemma \ref{PP-2} and define  
$r_o=\max\{r\in [0,R];u([0,r))<0\}$. 
 In the case that $r_o=R$,  using (\ref{pendent1}) and (\ref{P-2}) we have 
$$\cosh\psi(R)-1<\frac{\kappa}{4}(u(R)-u_0^2)<0.$$
Thus, $\cosh\psi(R)>1$ and consequently $\sinh\psi(R)>0$: the proof of Lemma \ref{PP-2} would says us that  $R$ is 
not  maximal. Therefore, $r_o<R$ and $u(r_o)=0$. This means that $r_o$ is the first zero of $u$ and that 
the function $u$ increases until a value $R$ where $R>r_o$. 
Let us see now that there will be a first time where $u$ attains a maximum, and next, $u$ decreases.

\begin{lemma} \label{p-concave}
In the above conditions, let $r_a<r_b$ be two numbers such that 
$u(r_a)<0<u(r_b)$ and $|u(r_a)|=u(r_b)$. Then $u'(r_b)<u'(r_a)$. Furthermore, in the interval $[0,R]$ $u$ is concave provided $u>0$.
\end{lemma}

\begin{proof} We integrate the expression  (\ref{pendent1}) from $u=u(r_a)$ to $u=u(r_b)$ obtaining
$$\cosh\psi(r_b)-\cosh\psi(r_a)=\kappa\frac{u(r_b)^2-u(r_a)^2}{2}-\int_{u_a}^{u_b}
\frac{\sinh\psi}{r}\ du=-\int_{u_a}^{u_b}
\frac{\sinh\psi}{r}\ du<0.$$
Because $\psi>0$, this implies $\psi_b<\psi_a$ and so, 
$$u'(r_b)=\tanh(\psi(r_b))<
\tanh(\psi(r_a))=u'(r_a).$$
Suppose now that $u>0$. Then (\ref{pendent11}) becomes
\begin{equation}\label{concave}
\frac{u''}{(1-u'^2)^{3/2}}=(\sinh\psi)'=\kappa u-\frac{\sinh\psi}{r}<\frac{\kappa u}{2}<0,
\end{equation}
where we have used (\ref{P-2}). Thus $u''<0$.

\hfill{$q.e.d$}
\end{proof}
As conclusion, as we take values beyond the point $r_o$, the function is concave, and because $u'\rightarrow 0$ as 
$r\rightarrow 0$, Lemma \ref{p-concave} implies that there will be a first point $r=r_M$ where 
the function $u$  presents a strict local maximum. Thus the function  $u$ is strictly increasing in the  interval $(0,r_M)$, across the 
$r$-axis in some point $r_o$  and it continues until a strict local maximum $r=r_M$, where 
the function $u$ attains  a height $u_M$, with 
$u_M<|u_0|$. Moreover,  in the interval $[r_o,r_M]$ the function $u$ is concave  and no inflection points exist in 
$[r_o,r_M]$.

It can be precised  the difference value between $u_M$ and $u_0$ with the following 

\begin{lemma} \label{pendent-lemma}Let $u$ be a solution of (\ref{rotational3})-(\ref{rotational33}). 
Denote $u_M$ the value of the first maximum of $u$, with $u(r_M)=u_M$. 
If  $\psi_o$ denotes the value of $\psi$ where $u=0$, then
\begin{equation}\label{pendent2}
u_M^2<\frac{2}{-\kappa}(\cosh\psi_o-1)<\frac{u_0^2}{2}.
\end{equation}
\end{lemma}

\begin{proof}
Integrating in (\ref{pendent1}) from $u=0$ to $u=u_M$ and taking into account that
 $u'(\psi_M)=0$ and $\sinh\psi_M=0$, we have,
$$\int_0^{u_M}\frac{\sinh\psi}{r}\ du+1-\cosh\psi_o=\kappa\frac{u_M^2}{2}.$$
But the first summand is positive, obtaining the first  inequality in (\ref{pendent2}). By using (\ref{P-2})
$$\int_{u_0}^0\frac{\sinh\psi}{r}\ du>\int_{u_0}^0\frac{\kappa u}{2}\ du=-\frac{\kappa u_0^2}{4}.$$
An integration of (\ref{pendent1}) from $u=u_0$ to $u=0$ leads to 
$$\int_{u_0}^0\frac{\sinh\psi}{r}\ du+\cosh\psi_o-1=-\frac{\kappa u_0^2}{2},$$
that together the last inequality gives the other inequality in (\ref{pendent2}).

\hfill{$q.e.d$}
\end{proof}

We can carry the same reasoning as in  Lemma \ref{pendent-lemma}  for the 
behaviour of $u$ as $r\rightarrow\infty$.  Then it is proved that 
 the function $u$ goes attained maximum and minimum points and that 
between two extrema there is exactly one zero of $u$.  

In the same sense, we obtain a generalization of (\ref{P-2})

\begin{proposition} Let $u$ be a solution of (\ref{rotational3})-(\ref{rotational33}). 
Denote by $m$ and $M$ a minimum and maximum point respectively for the function $u$.
Then there holds between any extremum and the following zero 
$$\frac{\kappa u}{2}<\frac{r}{r^2-r_m^2}\sinh\psi<\frac{\kappa u_m}{2}$$
$$\frac{\kappa u_M}{2}<\frac{r}{r^2-r_M^2}\sinh\psi<\frac{\kappa u}{2}$$
while  between any zero and the following extremum there holds 
$$-\frac{\kappa u_m}{2}<\frac{r}{r_m^2-r^2}\sinh\psi<-\frac{\kappa u}{2}$$
$$-\frac{\kappa u}{2}<\frac{r}{r_M^2-r^2}\sinh\psi<-\frac{\kappa u_M}{2}$$
\end{proposition}

\begin{proof} We integrate (\ref{pendent11}) between an extremum and we bound $u(t)$ in 
the integrand by the values  of $u$ at the correspondent extremum. Moreover, we use 
the fact that $\sinh\psi(r_m)=\sinh\psi(r_M)=0$. 

\hfill{$q.e.d$}
\end{proof}

In the same way, we obtain a similar result as in Lemma \ref{pendent-lemma}. We omit the details of the proof.

\begin{proposition} Let $u$ be a solution of (\ref{rotational3})-(\ref{rotational33}). Then we have the following:
\begin{enumerate}
\item Let $r_M$ and $r_m$ be two values of $r$ where $u$ attains two consecutive extrema. If 
 $r_o$ denotes the unique zero of $u$ in $(r_M,r_m)$, then
$$u_m^2<\frac{r_o^2+r_M^2}{2 r_o^2} u_M^2.$$
\item Let $r_m$ and $r_M$ be two values of $r$ where $u$ attains two consecutive extremal points. If 
 $r_o$ denotes the unique zero of $u$ in $(r_m,r_M)$, then
$$u_M^2<\frac{r_o^2+r_m^2}{2 r_o^2}u_m^2.$$
\end{enumerate}
\end{proposition}

As conclusion of the above results, the values of $u$ in these extrema goes decreasing to 
zero as $r$ goes to infinity, and the same occurs for the function $u$. 
We summarize the results obtained as follows:

\begin{theorem} Let $u$ be a solution of (\ref{rotational3})-(\ref{rotational33}). Then there holds the 
following:
\begin{enumerate}
\item The function $u$ has an infinity of zeros.
\item The function $u$ goes to $0$ as $r\rightarrow\infty$.
\item Between two successive extrema occurs exactly one inflection point. These points lie in the 
interval between a minimum and the following zero of $u$, and between a maximum and the following zero of $u$.

\end{enumerate}
\end{theorem}

\begin{proof}
We have only to show the item 3. Since the reasoning repeats in each interval of two successive extrema of $u$, we do 
 it only in the first interval $[0,r_M]$. Consider $r_o$ the first zero of $u$. It follows from Lemma \ref{p-concave}
that  $u''<0$ in $[r_o,r_M]$.
So that we have  to show that there exists exactly one inflection  in $(0,r_o)$. Denote $v=u'/\sqrt{1-u'^2}$ as in 
Theorem \ref{t-exi}. We know
that 
$$v'+\frac{v}{r}=\kappa u,\hspace*{.5cm}v(0)=0,\ \ v'(0)=-u_0/2.$$
 This means that $v$ is increasing on the right of $r=0$. Since 
$v(r_M)=0$, there exists one local maximum of $v$ in $(0,r_M)$, that is, a point of inflection. Let $r_1<r_o$ the first 
maximum of $v$, and so, an inflection point. In the next inflection $r=r_2$, $r_2<r_o$, we know that $v'(r_2)=0$ and 
$v''(r_2)\geq 0$. However, as $v'(r_2)=0$, then (cf. Eq. (\ref{eq-v}))
$$\frac{v(r_2)}{r_2}=\kappa u(r_2)>0.$$
Since $v''+(v/r)'=\kappa u'$, then at the point $r=r_2$ we have 
$$\kappa u'(r_2)\geq\left(\frac{v}{r}\right)'(r_2)=0,$$
which is a contradiction since $u'>0$ in $(0,r_M)$. In particular, the (infinite) set of inflection points of $u$ is isolated, and thus, this set as well as the set of the zeroes of $u$ is not bounded.

\hfill{$q.e.d.$}
\end{proof}

Our original problem of the existence of pendent liquid  drops can be solved. We consider that the radius and the 
angle of contact, or the support plane and the angle of contact are prescribed. The study is concentrated
in pendent liquid drops physically realizable, that is, drops that lie in one side of the support plane.
For the proof we need the following

\begin{lemma} \label{pendent-aux} 
Consider $u$ a solution of (\ref{rotational3})-(\ref{rotational33}) and 
denote $r_o$ the first zero of $u$. Then 
\begin{equation}\label{pendent-zero}
r_o>\sqrt{u_0^2-\frac{4}{\kappa}}>\frac{2}{\sqrt{-\kappa}}.
\end{equation}
\end{lemma}

\begin{figure}[htbp]
\includegraphics[width=9.8cm,height=5.5cm]{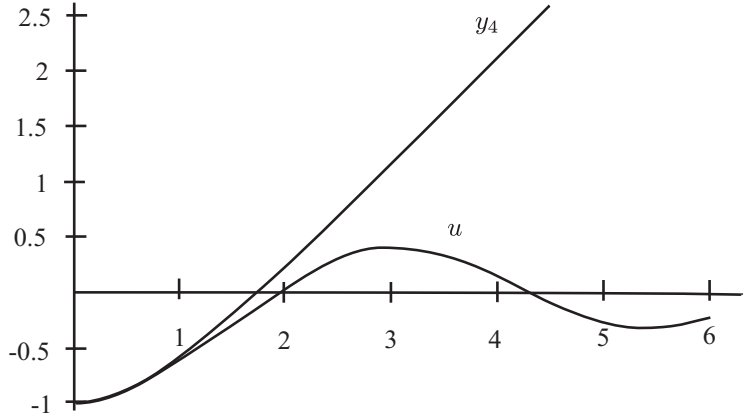}
\caption{Comparison between $y_4$ and $u$. Here $\kappa=-2$, $u_0=-1$.}\label{comparison}
\end{figure}

\begin{proof}
We consider the hyperbolic plane $\Sigma^4$ obtained by rotating the curve 
$$y_4(r;u_0)=\sqrt{r^2+\left(\frac{2}{\kappa u_0}\right)^2}+u_0-\frac{2}{\kappa u_0}$$
with respect to the $x_3$-axis. We prove that $y_4(r)>u(r)$. The derivative of $u$ is 
$$u'(r)=\tanh\psi(r)=\frac{\sinh\psi}{\sqrt{1+\sinh^2\psi}}=\frac{r}{\sqrt{r^2+r^2/\sinh^2\psi}}.$$
Using (\ref{P-2}), $y_4'>u'$ and because $y_4$ and 
$u$ have the same  initial conditions, then $y_4>u$. See Figure \ref{comparison}.
  Since $y_4$ meets the $r$-axis at the point
$\sqrt{u_0^2-4/\kappa}$, a comparison between $y_4$ and $u$  gives the desired estimate.

\hfill{$q.e.d$}
\end{proof}

\begin{theorem}[Existence of pendent liquid drops] \label{pendent-ex}
Let $\l^3$ be the Lorentz-Minkowski space. Let $\kappa<0$ be a constant of capillarity, and $\beta,\lambda\in \r$.
\begin{enumerate} 
\item Given $R>0$, there exists a stationary pendent liquid drop $X$ supported in a disc of radius $R>0$ in 
some horizontal spacelike plane $\Pi$ and such that $\beta$ is the contact angle along $\partial(\overline{X}\cap \Pi)$.
The mean curvature of a point $x\in X$ is 
$\kappa x_3(x)+\lambda$. The drop lies in one side of $\Pi$.
\item Consider  $\Pi$ a  spacelike plane. Then there exists a stationary pendent liquid drop $X$ supported on $\Pi$ that makes  
a constant angle $\beta$ along $\partial(\overline{X}\cap \Pi)$. The mean curvature of a point $x\in X$ is 
$\kappa x_3(x)+\lambda$, where $x_3$ denotes the distance to $\Pi$. The drop lies in one side of $\Pi$.

\end{enumerate}
\end{theorem}

\begin{proof}  We may assume without loss of generality that  $\lambda=0$ and that in the 
item 2, the plane $\Pi$ is horizontal. We know that the solution is obtained by rotating a solution 
$u$ of (\ref{rotational3})-(\ref{rotational33}) by an appropriate choice of 
$u_0$. Moreover, we can suppose that  $\beta\geq 0$ and $u_0\leq 0$. 

\begin{enumerate}
\item If $\beta=0$, the solution is $u=0$.
Let $\beta>0$. We will seek a solution $u$ such that $u'(R)=\tanh\beta$. 
Take $u_0$ sufficiently big so that $R< \sqrt{u_0^2-4/\kappa}$. We know that the hyperbolic cap $y_4$ satisfies $y_4>u$.
Furthermore, $u(t)<y_4(t)<0$ for $t\in [0,R]$. Then 
$$\sinh\psi(R)=\frac{\kappa}{R}\int_0^R t u(t;u_0)\ dt>\frac{\kappa}{R}\int_0^R t y_4(t;u_0)\ dt.$$
A direct integration gives that 
$$\lim_{u_0\rightarrow-\infty}\frac{\kappa}{R}\int_0^R t y_4(t;u_0)\ dt=+\infty.$$ 
Thus $u'(R;u_0)=\tanh\psi(R;u_0)\rightarrow 1$ and hence, take $|u_0|$ sufficiently big so the corresponding solution $u=u(r;u_0)$ of (\ref{rotational3})-(\ref{rotational33}) 
 satisfies $u'(R;u_0)>\tanh\beta$. Now, we take $u_0\rightarrow 0$. Since $u(r;u_0)\rightarrow 
0$ if $u_0\rightarrow 0$, the continuity on the parameter $u_0$ leads to the 
existence of  a number $u_0$ such that $u'(R;u_0)=\tanh\beta$.

\item Suppose $\Pi\equiv x_3=0$. If $\beta=0$, the solution is $u=0$.
Let $\beta>0$. Denote $r_o(u_0)$ the first zero of $u$  and let us study 
$u'(r_o;u_0)$ for each solution $u=u(r;u_0)$ where $u_0$ varies from $-\infty$ until $0$. 
By the continuity of the solutions with respect to the parameter $u_0$ and by Lemma \ref{pendent-aux}
$$\lim_{u_0\rightarrow 0}r_o(u_0)=0,\hspace*{1cm}\lim_{u_0\rightarrow-\infty}r_o(u_0)=\infty.$$
Furthermore, and since the hyperbolic plane $y_4$ lies over $u$, we have from (\ref{pendent1})
$$\sinh\psi(r_o)=\frac{\kappa}{r_o}\int_0^{r_o}t u(t)\ dt>\frac{\kappa}{s_o}\int_0^{s_o}
t y_4(t)\ dt\rightarrow \infty,$$
where $s_o$ is the (unique) zero of $y_4$ with the $r$-axis. Thus 
$u'(r_o;u_0)$ varies from $0$ to $1$ as $u_0$ goes from to $0$ to $-\infty$.
\end{enumerate}

\hfill{$q.e.d.$}
\end{proof}

Also we have a control of the first maximum $u_M$ of $u$ with its dependence on $u_0$.

\begin{corollary} Let $u_M=u_M(u_0)$ be the first maximum of $u=u(r;u_0)$, where $u$ is a solution 
of (\ref{rotational3})-(\ref{rotational33}), with $u_0<0$. Then 
$$\lim_{u_0\rightarrow -\infty}u_M=\infty.$$
In particular, the height $q=u_M-u_0$ increases monotonically with $u_0$.
 \end{corollary}

\begin{proof} Denote $r_o$ the first zero of $u$. By using  (\ref{pendent1}) and (\ref{P-2}), we have
$$\kappa\frac{{u_M}^2}{2}=\int_0^{u_M}\frac{\sinh\psi}{t}\ du+1-\cosh\psi(r_o)<1-\cosh\psi(r_o)+\kappa\frac{u_0 u_M}{2}.$$
By Lemma \ref{pendent-lemma}, 
$$\kappa\frac{u_0 u_M}{2}>\cosh\psi(r_o)-1+\kappa\frac{{u_0}^2}{4}.$$
Again, (\ref{P-2}) leads to $\sinh\psi(r_o)>\kappa u_0 r_o/2$, and then
$$\kappa\frac{u_0 u_M}{2}>\frac{\frac{\kappa^2 {u_0}^4}{4}}{1+\sqrt{1+\frac{\kappa^2 {u_0}^4}{4}}}+\frac{\kappa {u_0}^2}{4}.$$
From here, 
$$u_M>\frac{u_0}{2}\left(1+\frac{\kappa {u_0}^2}{1+\sqrt{1+\frac{\kappa^2 {u_0}^4}{4}}}\right)\longrightarrow\infty,$$
as $u_0\rightarrow-\infty$.

\hfill{$q.e.d.$}
\end{proof}

As conclusion of the proof of Theorem \ref{pendent-ex}, the solution $u(r;u_0)$ tends uniformly to a hyperbolic plane  as $u_0\rightarrow-\infty$. Exactly, 

\begin{corollary} Fix $\kappa<0$ and let $R>0$. Then for each $\epsilon>0$, there exists a constant 
$M=M(R,\epsilon)$ such that for any $u_0<0$, with $|u_0|>M$, it holds 
$$|y^{n)}_4(r;u_0)-u^{n)}(r;u_0)|<\epsilon,\hspace*{1cm}0\leq r\leq R, \ n\in\{0,1,2\}$$
\end{corollary}

\begin{proof} If we fix $R>0$, we know that 
$$y'_4(R;u_0)-u'(R;u_0)=\frac{R}{\sqrt{R^2+4/(\kappa u_0)^2}}-
\frac{R^2}{\sqrt{R^2+R^2/\sinh^2\psi(R)}}\longrightarrow 0,$$
if $u_0\rightarrow -\infty$, since $\sinh\psi(R;u_0)\rightarrow\infty$. The 
same occurs for $0\leq r\leq R$ since both $y_4$ and $u$ are increasing in the interval $[0,R]$.
Because both $y_4$ and $u$ agree at $r=0$, then $|y_4(R;u_0)-u(R;u_0)|\rightarrow 0$ as 
$u_0\rightarrow-\infty$. Finally, by using (\ref{rotational3}) and (\ref{P-2})
$$\frac{u'}{\sqrt{1-u'^2}}+r\frac{u''}{(1-u'^2)^{3/2}}=\kappa r u<
2\frac{u'}{\sqrt{1-u'^2}}.$$
  If we fix $R$, and for $|u_0|$ sufficiently big, 
we know that $R$ lies on the left of the first inflection point of $u(r;u_0)$, and so, $u$ is 
convex in $(0,R]$. Thus, for a fix $r\in (0,R]$, 
$0< r u''(r)<u'(r)(1-u'(r)^2)\rightarrow 0$
as $u_0\rightarrow -\infty$, since $u'(r)\rightarrow 1$. A direct computation leads to $\lim_{u_0\rightarrow-\infty}y_4''(r)=0$ again.

\hfill{$q.e.d.$}
\end{proof}

We want to estimate the enclosed volume by $u$ with the support plane. 
In general, the volume of a pendent liquid drop supported on a spacelike plane $\Pi$  is given as follows.
Without loss of generality, we assume $\Pi$ is parallel to $x_3=0$ and that $\lambda=0$ in (\ref{rotational1}). 
Thus the drop is obtained by a profile curve $u$ that is solution of (\ref{rotational3})-(\ref{rotational33}).
Let $r>0$ and $\psi$ the hyperbolic angle of contact between the graph of $u$ and the plane $\Pi\equiv
\{x_3=u(r)\}$. We consider the portion of $u$ between $r=0$ and $r=r_M$, where 
$r_M$ is the first maximum of $u$. 
Denote by ${\cal V}={\cal V}(r)$ the volume of  the drop 
determined by $u$  in the interval $[0,r]$. Setting $r_o$ 
the first zero of $u$, we have 
\begin{enumerate}
\item If $r\leq r_o$, then 
\begin{equation}\label{vo1}
{\cal V}(r)=-\frac{2\pi r \sinh\psi}{\kappa}
\end{equation}
\item If $r_o<r\leq r_M$, then 
\begin{equation}\label{vo2}
{\cal V}(r)=\pi r^2 u(r)-\frac{2\pi r \sinh\psi}{\kappa}.
\end{equation}
\end{enumerate}

\begin{figure}[htbp]
\includegraphics[width=9.4cm,height=3.5cm]{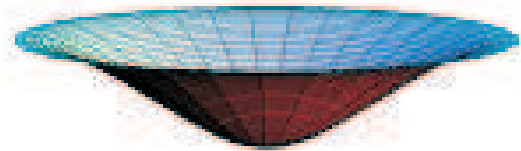}
\caption{Maximum pendent liquid drop. Here $\kappa=-2$, $u_0=-1$.}\label{maximum2}
\end{figure}

We define the {\it maximum volume} of $u$ as the 
volume determined by $u$ in the interval $[0,r_M]$, and  {\it maximum drop} the 
corresponding surface. This  has a physical significance since 
 any drop that extends beyond the coordinate $r_M$ would have 
to pass through again the supporting plane, which is physically unrealistic. See Figure \ref{maximum2}. 
We are going  to estimate a lower bound 
for ${\cal V}_M$. For $u$, we know that 
${\cal V}_M=\pi r_M^2 u_M$. 

On the other hand, the function $y_4$ lies above $u$ and this gives an estimate for 
${\cal V}$. The function $y_4$ attains the value $u_M$ at 
$$r_1=\sqrt{q^2+\frac{4}{\kappa u_0}q},$$
where $q=u_M-u_0$ is the height of the liquid drop.
Then the volume $V_M$ of $\Sigma^4$ in the interval $[0,r_1]$ is less than ${\cal V}_M$. In the 
same way, the volume ${\cal V}_0$ determined by $u$ with the $r$-axis can be estimated by below by 
the corresponding volume $V_0$ of $y_4$.  
The values of the volume of $\Sigma^4$ until the $r$ axis and until $y_4=u_M$ are respectively
$$V_0=\frac{\pi u_0}{3\kappa}(6-\kappa u_0^2),\hspace*{1cm}
V_M=\frac{\pi}{3\kappa u_0}q^2(6+\kappa u_0 q).$$

\begin{corollary} Let $u=u(r;u_0)$ be a solution of (\ref{rotational3})-(\ref{rotational33}). Then the maximum volume ${\cal V}_M$ and the volume ${\cal V}_o$  that makes the drop  with the planes $x_3=0$ and $x_3=u_M$ satisfy respectively
\begin{equation}\label{pendent-vo1}
\frac{\pi}{3\kappa u_0}q^2(6+\kappa u_0 q)<{\cal V}_M.
\end{equation}
\begin{equation}\label{pendent-vo2}
\frac{\pi u_0}{3\kappa}(6-\kappa u_0^2)<{\cal V}_0.
\end{equation}
\end{corollary}

Since $q=u_M-u_0\rightarrow \infty$ as $u_0\rightarrow-\infty$ (or ${\cal V}>{\cal V}_o\rightarrow\infty$), we conclude

\begin{corollary} In Lorentz-Minkowski space $\l^3$ there are realistic and physically realizable  pendent liquid drops 
with arbitrary big volume. Moreover, the drop is asymptotic to a hyperbolic cap.
\end{corollary}

We compare with Euclidean ambient: in Euclidean 3-space $\r^3$,  consider a small pendent liquid drop supported in a plane. 
When we  add more a more liquid in the drop, there exists a time where the drop turns into a state of instability  in such way that the drop falls down. 
In contrast to this, the above Corollary roughly says  that in Lorentz-Minkowski space, we can increase the amount of liquid in the pendent drop so that 
the drop increases its size, but the drop will never  fall down.
In the same way, the height of the maximum drop increases monotonically with volume as well as the
area of contact with the horizontal plate.

\begin{corollary} Let $u$ be a solution of (\ref{rotational3})-(\ref{rotational33}). Denote $r_o$ and $r_M$ the 
first zero and maximum of $u$. Then
$$\frac{-u_0(6-\kappa u_0^2)}{6\sinh\beta_0}<r_o,\hspace*{1cm}\sqrt{\frac{q^2(6+\kappa u_0 q)}{3\kappa u_0 u_M}}<r_M,$$
where $\beta_0$ is the hyperbolic angle that makes $u$ with the $r$-axis.
\end{corollary}

We prove now  the existence of pendent liquid drops for prescribed volume and angle of contact (see Theorem \ref{sessile-vo} for 
the case $\kappa>0$). 

\begin{theorem}\label{pendent-vo} Let $\kappa<0$ be  a constant of capillarity . 
Then for each  $V>0$ and $\beta\in\r$, there exists a stationary
pendent liquid drop $X$ in Minkowski space $\l^3$ supported in a spacelike plane  $\Pi$ such that 
the volume of the drop is $V$  and $\beta$ is the boundary angle of contact between $X$ 
and $\Pi$ along $\partial (\overline{X}\cap\Pi)$. This pendent liquid drop lies in one side of $\Pi$.
\end{theorem}

\begin{proof} Without loss of generality, we suppose that $\beta>0$ and that the drop $X$ is 
defined by a solution $u$ of (\ref{rotational3})-(\ref{rotational33}) for $u_0<0$. Let $R$ be the real number such that 
$$V=-\frac{2\pi R\sinh\beta}{\kappa}.$$
If $R<R_o:=2/\sqrt{-\kappa}$, we consider the solution $u(r;u_0)$ such that the pendent drop 
that defines is supported in a horizontal plane $\Pi$ where  
$R$ is the radius of the disc that the drop wets on $\Pi$ and $\beta$ is the angle of contact. By Lemma 
\ref{pendent-aux}, $R$ lies on the left of the first zero $r_o$ of $u$. As a consequence, 
the volume ${\cal V}$ of the drop is exactly $V$ by (\ref{vo1}).  Thus, ${\cal V}$ takes all values in 
the interval $[0,-\frac{2\pi R_o\sinh\beta}{\kappa}]$.

Consider now $V>-\frac{2\pi R_0\sinh\beta}{\kappa}$ and let $r$ be such that 
$V=-\frac{2\pi r\sinh\beta}{\kappa}$ and the corresponding solution $u=u(r;u_0)$ of 
(\ref{rotational3})-(\ref{rotational33}) with $u'(r)=\tanh\beta$. If $r<r_o$, $r_o$ the first zero of $u$, then
we obtain ${\cal V}=V$ again. On the contrary case, by (\ref{vo2}), 
$${\cal V}(r;u_0)=\pi r^2 u(r,u_0)+V=\pi r^2 u(r)-\frac{2\pi r\sinh\beta}{\kappa}>-\frac{2\pi R_o\sinh\beta}{\kappa}.$$
Now, we solve  (\ref{rotational3})-(\ref{rotational33}) with $u'(r)=\tanh\beta$ and $r\searrow R_0$.
By the continuity of ${\cal V}$ with respect to $(r;u_0)$
 and since 
$${\cal V}(r;u_0)\leq \pi r_M^2 u_M={\cal V}_M\longrightarrow 0$$
 as $u_0\rightarrow 0$, ${\cal V}(r;u_0)$ takes any value until to arrive to $0$.

\hfill{$q.e.d$}
\end{proof}

\begin{remark} In Theorems \ref{pendent-vo} and \ref{pendent-ex}, we found a qualitative difference 
with the Euclidean case. Remember that in Euclidean space, for a given constant of capillarity, 
 there exists a universal finite upper bound for the physical diameter for all possible pendent drops \cite{cf}.
However, in Lorentzian setting, the  existence of pendent liquid drops when 
a volume and an angle of contact or when a disc of wetting and an angle of contact are prescribed is 
assured for any {\it arbitrary} volume or disc. 
 Furthermore, the solutions correspond with a  physical realistic pendent drop in the sense that the drop lies in one side of the supporting plane $\Pi$  and 
do not across $\Pi$ beyond the liquid-air-solid interface.
\end{remark}

\begin{remark}\label{pendent-mono} For pendent liquid drops we have not a result of monotonicity with respect to the constant of 
capillarity, as in Theorem \ref{sessile-mono}, even in the case that we only consider the profile curve until 
its first maximum. For example, if we take pendent drops until $r_M$, then 
the contact angle  is $0$, but the profiles do not present a monotony of shapes.  

However,  we have a similar inclusion property  as in  the Claim of Theorem \ref{foliation}. 
Consider    $u=u(r;u_0)$ the solution of 
(\ref{rotational3})-(\ref{rotational33}) with $u_0<0$. Denote  $r_M(u_0)$ the first maximum of $u$. Let $\delta>0$.
Then $u(r;u_0-\delta)>u(r;u_0)-\delta$ for each $r\in(0,r_M]$.
\end{remark}

%\begin{figure}[htbp]\label{nomonotony}
%\centereps{8.3cm}{5cm}{c:/articulos/sessile/mono.eps x=8.3cm y=5cm}
%\caption{Two pendent liquid drops for $\kappa=-1$ and $\kappa=-2$.}
%\end{figure}

%\begin{figure}[htbp]
%\centereps{8.12cm}{4.7cm}{c:/articulos/sessile/p2.eps x=8.12cm y=4.7cm}
%\caption{Profile of a maximum pendent liquid drop. Here $\kappa=-2$, $u_0=-1$.}
%\end{figure}

%%%%%%%%%%%%%%%%%%%%

\end{document}